%% file: main.tex
\documentclass[10pt, journal, comsoc]{IEEEtran}
\usepackage{acronym} \input{Utils/acronyms}
\usepackage{graphicx}
\graphicspath{ {./img/} }
\usepackage{cite}
\usepackage{array}
\usepackage{ifpdf}
\ifpdf
\else
\fi
\usepackage{amsmath}
\usepackage{bbm}
\usepackage{amssymb}
\usepackage[cmintegrals]{newtxmath}
\usepackage{enumitem}
\usepackage{booktabs}
\usepackage[normalem]{ulem}
\usepackage{url}
\usepackage{lscape}
\vspace{\baselineskip}
\usepackage{tabulary}
\usepackage{tabularx}
\usepackage{colortbl}
\usepackage{longtable}
\usepackage{multirow}
\usepackage{xcolor}
\usepackage{soul}
\usepackage[modulo]{lineno}
\usepackage{amssymb}
\usepackage{placeins}
\usepackage{float}
\usepackage{scalefnt}
\usepackage{multicol, blindtext, graphicx}
\useunder{\uline}{\ul}{}
\usepackage[utf8]{inputenc}
\usepackage{pifont}
\usepackage{lipsum}
\usepackage{subcaption}
\usepackage{units}
\usepackage[ruled,noend,linesnumbered]{algorithm2e} % For writing algorithms

\newcommand{\1}{(\textit{i})}
\newcommand{\2}{(\textit{ii})}
\newcommand{\3}{(\textit{iii})}

\DeclareMathOperator*{\maximize}{max}
\DeclareMathOperator*{\minimize}{min}

\newcommand{\yale}[1]{``#1''}

\begin{document}
% Control the number of names in references
\bstctlcite{IEEEexample:BSTcontrol}

%\title{Uncertainty-Aware Adaptive Reinstantiation of RAN Functions for Resilient 6G Networks}

\title{
\begin{flushleft} \normalfont\small This work has been submitted to the IEEE for possible publication.\\ Copyright may be transferred without notice, after which this version may no longer be accessible. \end{flushleft}
\hrule
\vspace{1em}
\Huge
Resilience under Uncertainty: Securing 6G through Stochastic Reinstantiation of RAN Functions}% to  Networks}

\author{

	\IEEEauthorblockN{
      Gabriel M. Almeida\IEEEauthorrefmark{1}\IEEEauthorrefmark{2},
      Gabriel Vieira\IEEEauthorrefmark{2},
      Jacek Kibi\l{}da\IEEEauthorrefmark{2},
      Joao F. Santos\IEEEauthorrefmark{2}, 
      Kleber Vieira Cardoso\IEEEauthorrefmark{1}
    }
    
	\IEEEauthorblockA{
      \IEEEauthorrefmark{1}\textit{Universidade Federal de Goiás}, Brazil, e-mails: gabrielmatheus@inf.ufg.br, kleber@ufg.br\\
      \IEEEauthorrefmark{2}\textit{Commonwealth Cyber Initiative}, \textit{Virginia Tech}, USA, 	e-mails: \{gabrielvieira, jkibilda, joaosantos\}@vt.edu	
    }
}

\maketitle

\IEEEpeerreviewmaketitle
\begin{abstract}
    \input{Sections/abstract.tex}
\end{abstract}


\acresetall

\input{Sections/1-introduction}
\input{Sections/2-background}
\input{Sections/3-related_work}
\input{Sections/4-system_model}
\input{Sections/5-problem_formulation}
\input{Sections/6-solution}
\input{Sections/7-evaluation}
\input{Sections/8-conclusion}
\input{Sections/acknowledgment}

%Referências Bibliográfica
\bibliographystyle{IEEEtran}
\bibliography{biblio}

\end{document}

%% file: Utils/acronyms.tex
\acrodef{5G}[5G]{Fifth Generation of Mobile Networks}
\acrodef{QoS}[QoS]{Quality of Service}
\acrodef{QoE}[QoE]{Quality of Experience}
\acrodef{AI}[AI]{Artificial Intelligence}
\acrodef{BS}[BS]{Base Station}
\acrodef{BS}[BS]{Base Station}
\acrodef{CN}[CN]{Core Network}
\acrodef{PN}[PN]{Processing Node}
\acrodef{NFV}[NFV]{Network Function Virtualization}
\acrodef{SDN}[SDN]{Software-Defined Networking}
\acrodef{RAN}[RAN]{Radio Access Network}
\acrodef{vRAN}[vRAN]{Virtualized \ac{RAN}}
\acrodef{MNO}[MNO]{Mobile Network Operator}
\acrodef{CU}[CU]{Centralized Unit}
\acrodef{DU}[DU]{Distributed Unit}
\acrodef{vCU}[CU]{Virtualized Central Unit}
\acrodef{vDU}[DU]{Virtualized Distributed Unit}
\acrodef{RU}[RU]{Radio Unit}
\newacroindefinite{RU}{an}{a}
\acrodef{O-RAN}[O-RAN]{Open \ac{RAN}}
\acrodef{ILP}[ILP]{Integer Linear Program}
\acrodef{TTI}[TTI]{Transmission Time Interval}
\acrodef{URLLC}[URLLC]{Ultra-Reliable Low-Latency Communications}
\acrodef{SINR}[SINR]{Signal-to-Interference-plus-Noise Ratio}
\acrodef{CoMP}[CoMP]{Coordinated Multi-Point}
\acrodef{SAA}[SAA]{Sample Average Approximation}
\acrodef{BTSP}[BTSP]{Bi-Parameterized Two-stage Stochastic Program}
\acrodef{HPPP}[HPPP]{Homogeneous Poisson Point Process}
\acrodef{PRB}[PRB]{Physical Resource Block}
\acrodef{SMIP}[SMIP]{Stochastic Mixed-Integer Programming}
\acrodef{PDF}[PDF]{Probability Density Function}
\acrodef{DFR}[DFR]{Deterministic Function Reallocation}
\acrodef{CQI}[CQI]{Channel Quality Indicator}
\acrodef{UMi}[UMi]{Urban Micro}
\acrodef{UMa}[UMa]{Urban Macro}
\acrodef{RMa}[RMa]{Rural Macro}
\acrodef{WS}[WS]{Wait-and-See}
\acrodef{CCDF}[CCDF]{Complementary Cumulative Distribution Function}

%% file: Sections/abstract.tex
The disaggregation of base stations into discrete \ac{RAN} functions introduces new threats to mobile networks, as failures in one \ac{RAN} function can trigger cascading failures and interrupt entire function chains, with potential to degrade network performance and disrupt service.
In this paper, we propose the first resilience mechanism for disaggregated mobile networks that leverages the adaptive reinstantiation of \ac{RAN} functions under uncertainty to mitigate disruptions and maintain service continuity in the presence of compromised infrastructure.
Our mechanism reacts to cascading failures that disrupt \acp{RU} by reinstantiating \acp{CU} and \acp{DU} in alternative cloud locations, restoring their function chains while accounting for uncertainty in users' locations and wireless channel conditions during the \textit{in-failure} state.
We formulate this recovery process as a two-stage stochastic optimization problem, where reinstantiation and routing decisions are made under uncertainty, and bandwidth allocation decisions are performed after uncertainty is resolved.
We solve the problem using a \ac{SAA}-based solution as a tractable, deterministic equivalent problem.
We numerically evaluate our approach on a real-world disaggregated mobile network topology across multiple failure scenarios and traffic demand conditions,  and our results demonstrate that our solution can achieve up to 80\% higher recovery performance compared to conventional resilience mechanisms.

%% file: Sections/1-introduction.tex
\section{Introduction}
\label{sec:introduction}

Disaggregation of mobile networks is a key architectural evolution toward 6G, which decomposes monolithic base stations into \emph{function chains} of \acp{CU}, \acp{DU}, and \acp{RU}, each implementing different layers of the \ac{RAN} protocol stack~\cite{santos2025managing}.
In the disaggregated mobile network architecture, \acp{RU} are deployed at cell sites to provide wireless connectivity to users, while \acp{CU} and \acp{DU} are softwarized and deployed at distributed \emph{cloud locations}, enabling mobile carriers to flexibly deploy and scale \ac{RAN} functions to achieve distinct objectives, e.g., optimizing costs and energy efficiency~\cite{morais:2023}.

Despite its operational benefits, the disaggregated mobile network architecture also increases the susceptibility of mobile networks to outages caused by failures, e.g., power outages~\cite{sai:2024} and conflicting control actions~\cite{zolghadr:25}.
Failures can compromise cloud locations hosting \ac{RAN} functions, interrupting the function chains of multiple \acp{RU} and triggering \emph{cascading failures}~\cite{Lazarev:2023}.
For example, a single compromised cloud location hosting a \ac{DU} interrupts the function chains of all its associated \acp{RU}, while a compromised cloud location hosting a \ac{CU} interrupts the function chains of all its associated \acp{DU}, and consequently their associated \acp{RU}.
As a result, \acp{RU} experiencing interrupted function chains are unable to provide wireless connectivity to users, disrupting service continuity and potentially leading to large-scale outages~\cite{almeida:2025}.

\begin{figure}[t]
  \centering
  \hfill
  \subfloat[a][Before failure.]{\includegraphics[width=0.1586\textwidth]{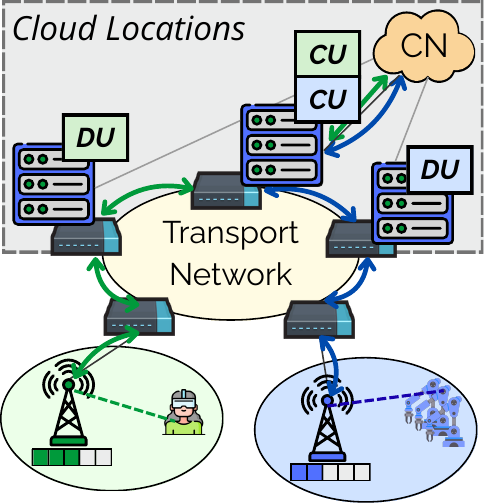}\label{fig:before_failure}}
  \hfill
  \subfloat[b][After failure.]{\includegraphics[width=0.1586\textwidth]{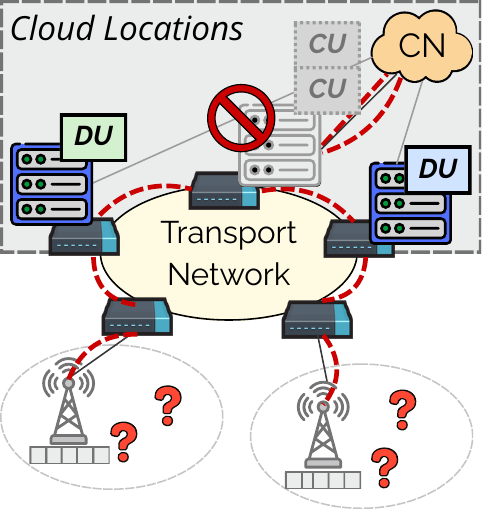} \label{fig:after_failure}}
  \hfill
  \subfloat[c][After recovery.]{\includegraphics[width=0.1586\textwidth]{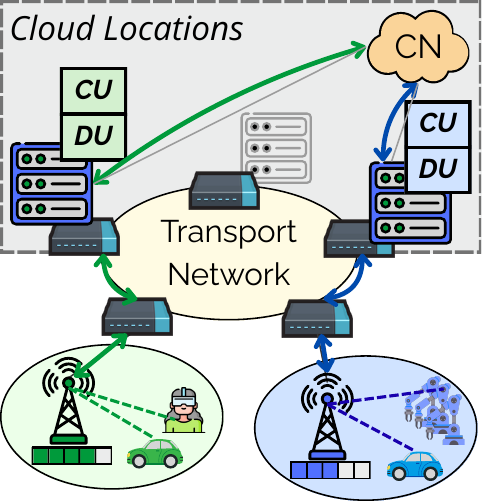}\label{fig:after_recovery}}
  \hfill
  %\vspace{+1em}
  \caption{Disaggregated mobile networks are susceptible to cascading failures, as disruption in a single \ac{RAN} function can compromise the entire function chain and cause outages.}
  \vspace{-1em}
  \label{fig:introduction}
\end{figure}

At the same time, the softwarized nature of disaggregated mobile networks introduces opportunities for \emph{resilience} -- understood as the ability to cope with failures and minimize the impact of disruptions to maintain service continuity~\cite{WalidSaad:2024}.
Softwarization enables the creation of adaptive strategies to respond to disruptions, e.g., by reinstantiating \ac{RAN} functions in alternative cloud locations to restore interrupted function chains and mitigate the impact of cascading failures, as illustrated in Fig.~\ref{fig:introduction}.
While there is a mature body of literature exploring disaggregated mobile network topologies and the placement of \ac{RAN} functions~\cite{Pires:2025,Kim:2025,You:2025,Hojeij:2025,Rocha:2023}, most of these works focus on improving network performance under normal operating conditions, without considering the impact of failures.

Several works investigate \emph{resilience mechanisms} in disaggregated mobile networks to mitigate the impact of failures~\cite{Scholler:2013,Carlinet:2019,Lazarev:2023,KAADA:24,comeback2023,almeida:24}.
These approaches typically rely on either redundancy-based strategies, e.g., deploying backup functions to recover from service disruptions~\cite{Scholler:2013,Carlinet:2019,Lazarev:2023}, or on adapting radio configurations considering the remaining operational \acp{RU}~\cite{comeback2023,KAADA:24,almeida:24}.
However, we observe a significant gap in the literature: existing resilience mechanisms in disaggregated mobile networks have not explored the restoration of interrupted function chains to recover the service of \acp{RU} after failures.

To address this gap, we propose the first adaptive resilience mechanism for disaggregated mobile networks that reinstantiates \ac{RAN} functions.
Our proposed mechanism builds on our previous conference paper~\cite{almeida:2025} by considering uncertainty in users' wireless channel conditions and locations arising from the fact that, after a disruption, \acp{RU} with interrupted function chains are unable to serve users or collect channel state information, e.g., \ac{CQI}.
In this paper, we propose a two-stage stochastic optimization framework that: 
\1 reacts to failures by reinstantiating \ac{RAN} functions to restore \acp{RU} function chains under uncertainty in users' wireless channel conditions and locations;
\2 restores service continuity by reassociating users after the \acp{RU} function chains are recovered; and
\3~achieves up to 80\% performance improvement over existing approaches, including our prior work~\cite{almeida:2025}, while\,increasing\,CPU usage by only 11\% on average.

Our key contributions can be summarized as follows:
\begin{itemize}
    \item We propose a resilience mechanism that restores interrupted function chains of \acp{RU} to recover service continuity by adaptively reinstating \ac{RAN} functions under uncertainty in users' locations and wireless channel conditions.
    \item We formalize the problem of determining \ac{RAN} function reinstantiation decisions under uncertainty and reassociating users after restoring the function chains of \acp{RU} as \iac{BTSP}.
    \item We propose a \ac{SAA}-based solution to solve the \ac{BTSP} problem, leveraging a deterministic approximation of the original problem.
    \item We evaluate our \ac{SAA}-based solution in a real-world mobile network topology, comparing its performance against state-of-the-art resilience mechanisms.
\end{itemize}

The remainder of this paper is organized as follows. 
In Section~\ref{sec:background}, we review background and define terminology.
In Section~\ref{sec:related_work}, we present the related works and highlight our contributions.
In Section~\ref{sec:system_and_formulation}, we define our system model of the disaggregated mobile network.
In Section~\ref{sec:problem_formulation}, we formulate our proposed two-stage stochastic optimization framework.
In Section~\ref{sec:solution}, we present our \ac{SAA}-based solution for solving the proposed two-stage stochastic optimization framework.
In Section~\ref{sec:evaluation}, we present numerical results validating our solution and assessing its performance across different failure scenarios in comparison with resilience mechanisms in the literature.
Finally, in Section~\ref{sec:conclusion}, we pose our concluding remarks and discuss potential directions for future work.

%% file: Sections/2-background.tex
\section{Technical Background}
\label{sec:background}
In this section, we present the definitions of robustness, reliability, and resilience, and discuss how resilience mechanisms can recover service over time.
Then, we introduce the terminology adopted throughout the rest of this paper.

\subsection{Robustness, Reliability, and Resilience}

In system design, \emph{resilience} is often considered alongside \emph{reliability} and \emph{robustness} as part of the R\textsuperscript{3} framework for designing systems that can \yale{withstand, absorb, adapt to, and
recover from disturbances}~\cite{zissis2019r3}.
While there is some overlap between the three concepts, their definitions differ fundamentally and lead to distinct designs.
Reliability quantifies the long-term probabilistic absence of failures, robustness quantifies the impact of anticipated failures, while resilience characterizes the system's ability to cope with unexpected failures.
The latter can be defined as the ability of a system to \textit{absorb}, \textit{adapt}, and \textit{recover} from unexpected failures, restoring service continuity while minimizing service degradation~\cite{WalidSaad:2024}.

\begin{figure}[t]
    \centering
    \includegraphics[width=1\linewidth]{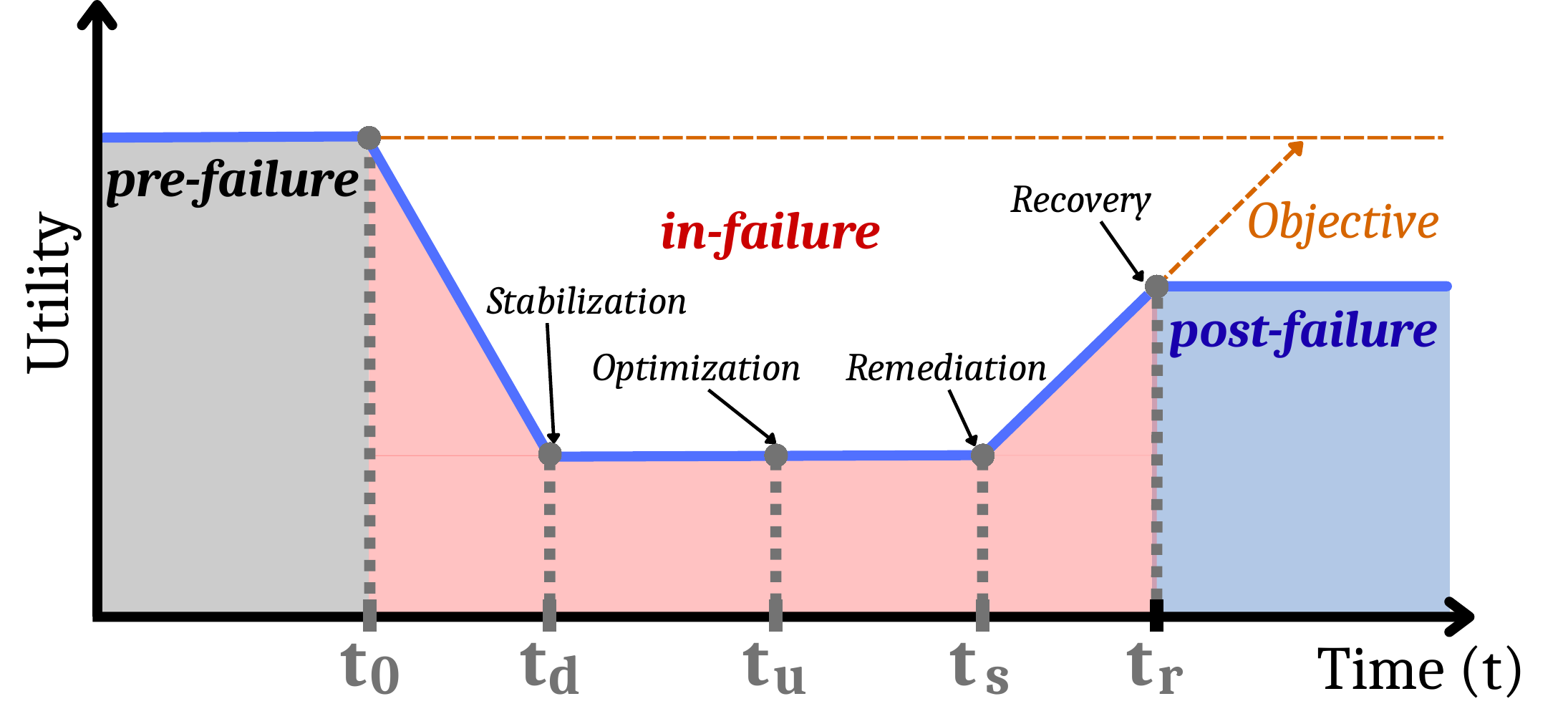}
    \caption{Illustrative communication resilience dynamics showing the pre-failure, in-failure, and post-failure states as reflected by the evolution of network utility over time.}
    \label{fig:temporal_evolution_illustration}
\end{figure}

The impact of failures in mobile networks and the effects of resilience mechanisms can be modeled through the temporal evolution of network utility, as proposed in~\cite{WalidSaad:2024}.
In this context, network utility captures the overall service level provided to users as the network transitions from normal operation to failure and subsequent recovery.
To characterize this evolution, we consider a sequence of five key events: \textit{failure} ($t_0$), \textit{stabilization} ($t_d$), \textit{optimization} ($t_u$), \textit{remediation} ($t_s$), and \textit{recovery} ($t_r$), each representing a distinct stage in the network response to failures.
Figure~\ref{fig:temporal_evolution_illustration} illustrates the temporal evolution of network utility under failures, highlighting these key events.

Initially, the network operates in the \textit{pre-failure} state, where all \acp{RU} and cloud locations are fully operational.
At time $t_0$, a failure occurs, transitioning the network to the \textit{in-failure} state.
As a consequence, network services are disrupted, leading to a degradation in network utility.
At time $t_d$, the network absorbs the impact of the failure, stabilizing the degraded utility.
At time $t_u$, the network detects the disruptions and triggers a resilience mechanism, initiating the process of defining recovery actions to restore service continuity.
At time $t_s$, recovery actions are determined, and the network begins to apply them, gradually restoring disrupted services and improving network utility.
Finally, at time $t_r$, the network reaches the \textit{post-failure} state, in which recovery actions are completed, and service is restored for previously disrupted \acp{RU} and disconnected users.

Although reliability and robustness are often considered in mobile network design, resilience remains a relatively unexplored territory. 
This is changing as part of 6G pre-standardization efforts, with resilience being incorporated into network design and operation to enable autonomous detection, response, and recovery from failures~\cite{nextgtrust,6gia}.

\subsection{Terminology}
\label{subsec:terminology}

To ensure consistency in describing failures and recovery actions, we adopt the following terminology in this paper:

\begin{itemize}
    \item \emph{Failure}: An internal or external event that disrupts the operation of network components (e.g., cloud locations).
    Examples include: conflicts~\cite{zolghadr:25}, misconfigurations~\cite{da20245grecon}, power outages~\cite{bi2024failure}, and extreme weather events~\cite{laurent:2022}.
    \item \emph{Compromised Cloud Location}: A cloud location affected by a failure and therefore unable to host \ac{RAN} functions.
    \item \emph{Interrupted Function Chain}: A function chain composed of a \ac{CU}, a \ac{DU}, and a \ac{RU} whose operation is disrupted due to compromised cloud locations hosting its functions.
    \item \emph{Disrupted \ac{RU}}: \Iac{RU} whose function chain is interrupted and cannot provide wireless connectivity to users.
    \item \emph{Operational \ac{RU}}: \Iac{RU} whose function chain remains intact and continues to provide wireless connectivity to users after a failure.
    \item \emph{Recovered \ac{RU}}: A previously disrupted \ac{RU} whose function chain has been restored through recovery actions, enabling it to resume providing wireless connectivity.
    \item \emph{Disconnected User}: A user that loses connectivity due to a disrupted \ac{RU} and is not served by any \ac{RU}.
    \item \emph{Non-affected User}: A user that remains associated with its original \ac{RU}, whose service is not affected by the failure.
    \item \emph{Re-associated User}: A disconnected user that is reassigned to \iac{RU} with available resources after a failure.
\end{itemize}

%% file: Sections/3-related_work.tex
\section{Related Work}
\label{sec:related_work}

In this section, we review the existing literature on \ac{RAN} function placement and resilience mechanisms,  and position our work against the state of the art.
%highlight how our resilience mechanism differs from existing approaches.

\subsection{Existing Approaches on \ac{RAN} Function Placement}

The placement of \ac{RAN} functions in disaggregated mobile networks can be formulated as a joint optimization problem that determines the functional split and the placement of softwarized \acp{CU} and \acp{DU} across distributed cloud locations, while the physical \acp{RU} remain deployed at cell sites~\cite{morais:2023}.
A substantial body of literature has investigated this problem, focusing on optimizing objectives such as energy consumption, operational costs, and latency requirements of network slices in disaggregated mobile networks.
For instance, \cite{Pires:2025}~proposed an optimization framework to maximize energy efficiency by allocating \ac{RAN} functions to cloud locations with lower power consumption.
Additionally, \cite{You:2025}~formulated a channel-aware placement algorithm to reduce energy consumption by jointly optimizing placement and the activation and deactivation of \acp{RU}.
The work of~\cite{Kim:2025} presented a data-driven model to minimize costs associated with the operation of \ac{RAN} functions, considering service interruptions during the reconfiguration of \ac{RAN} function placement.
Similarly, \cite{Rocha:2023}~presented a cost and latency-aware placement approach based on queuing theory for \ac{URLLC} applications.
Moreover, \cite{Hojeij:2025}~proposed a data-driven model to jointly optimize \ac{RAN} function placement and user association.

Existing approaches on \ac{RAN} function placement focus on optimizing network performance under normal operating conditions and do not account for failures in disaggregated mobile network infrastructures.
In particular, they assume that the underlying infrastructure remains operational, neglecting how failures may interrupt \acp{RU} function chains.
As a result, to the best of our knowledge, no prior work has investigated \ac{RAN} function placement as a resilience mechanism to support the reinstantiation of function chains after failures.

\subsection{Resilience Mechanisms for Mobile Networks}

Several works in the broader wireless networking literature propose resilience mechanisms based on redundancy to recover network service under anticipated failure scenarios.
For instance, \cite{Scholler:2013} presented a service blueprint that incorporates function replication to ensure service availability.
Similarly, \cite{Carlinet:2019} formulated an optimization problem that proactively replicates 5G network functions to recover from disruptions.
In \cite{Lazarev:2023}, the authors consider PHY-layer resilience by modeling failures in digital signal processing and recovering communication performance through replication of PHY \ac{RAN} functions.
While these works improve resilience to disruptions, they rely on replicating network functions or allocating redundant resources, implicitly assuming that sufficient infrastructure remains available after failures.
However, this assumption may not hold in the presence of unanticipated failures that compromise parts of the mobile network infrastructure~\cite{WalidSaad:2024}.

Only a limited number of works~\cite{Xing:2023, comeback2023, KAADA:24} have investigated resilience mechanisms in the context of unanticipated failures.
The work of~\cite{Xing:2023} explores migrating user traffic from disrupted \acp{DU} to operational \acp{DU} after failures, demonstrating the potential of dynamic user association to maintain service continuity.
However, this approach relies on the remaining operational \acp{DU} to restore service for disconnected users.
Similarly, \cite{comeback2023} introduced a three-layer resilience mechanism that jointly optimizes rate adaptation, beamforming, and user association in response to failures, relying on the remaining operational \acp{RU} to restore service continuity.
In addition, \cite{KAADA:24} focuses on adapting transmission power and antenna tilt to expand the coverage of operational \acp{RU} after failures and restore connectivity to users.
While these works explore different strategies to restore service using the available infrastructure after failures, they do not address the recovery of disrupted \ac{RAN} function chains, limiting their ability to restore service under large-scale disruptions caused by cascading failures.

In the conference version of this paper~\cite{almeida:2025}, we proposed a resilience mechanism based on \ac{RAN} function reallocation to restore service continuity by recovering interrupted function chains of disrupted \acp{RU} under cascading failures.
Our results demonstrated that restoring service for disrupted \acp{RU} provides greater network service recovery than approaches that rely solely on the remaining operational infrastructure after failures.
However, our initial approach assumes perfect knowledge of users' locations and wireless channel conditions after disruptions, which is unlikely (if not infeasible) in practice, as disrupted \acp{RU} are unable to collect information about users in their coverage areas, introducing uncertainty into the recovery process.
Addressing this limitation requires a completely new problem formulation and resilience mechanism. As such, this paper extends~\cite{almeida:2025} by proposing a resilience mechanism that reinstantiates \ac{RAN} functions to recover interrupted function chains of disrupted \acp{RU} under uncertainty in users' locations and wireless channel conditions. This leads to a novel problem formulation comprising a two-stage stochastic optimization framework that: \1 optimizes \ac{RAN} function reinstantiation decisions under uncertainty to mitigate cascading failures; and \2 allocates communication resources in the recovered \acp{RU} to restore service for disconnected users.

%% file: Sections/4-system_model.tex
\section{System Model}
\label{sec:system_and_formulation}

In this section, we present the system model of the disaggregated mobile network, describing its network topology, characterizing the impact of failures on its resources, and defining the sources of uncertainty associated with the recovery process. We denote sets by calligraphic letters (e.g., $\mathcal{R}$), decision variables by lowercase Latin letters with subscripts and superscripts (e.g., $x_r^p$), stochastic variables by uppercase Greek letters (e.g., $\Gamma_{u,r}$), and their realizations over a specific scenario $\omega \in \Omega$ by lowercase Greek letters (e.g., $\gamma_{u,r}^{(\omega)}$).
In addition, we denote resource capacities of cloud locations, transport links, and \acp{RU} by lowercase Latin functions (e.g., $d(n)$), while we denote requirements of users and \ac{RAN} functions by uppercase Latin letters (e.g., $C(r)$).
We explain additional notation as needed.

\subsection{Disaggregated Mobile Network}
\label{subsec:net-system}

We model the disaggregated mobile network coverage as a two-dimensional geographical region $\mathcal{A} \subset \mathbb{R}^2$, served by a set of \acp{RU} $\mathcal{R}$.
Each \ac{RU} $r \in \mathcal{R}$ is deployed at a fixed location $\phi_r\in\mathcal{A}$ and is assigned a finite bandwidth $b(r)$ and transmission power $p(r)$.
The \ac{CU} and \ac{DU} forming a function chain with each \ac{RU} $r \in \mathcal{R}$ are softwarized and hosted on a set of cloud locations $\mathcal{N}$, spread across the disaggregated mobile network topology.
Each cloud location $n \in \mathcal{N}$ has a processing capacity $d(n)$.
We consider the \acf{CN}, denoted by $v_0$, representing the source (downlink) and destination (uplink) for all the traffic to/from \acp{RU}.
Without loss of generality, we focus on the downlink case in this paper, as our system model and proposed resilience mechanism can be extended to the uplink case with minor modifications.

We model the disaggregated mobile network topology as a graph $\mathcal{G} = (\mathcal{V}, \mathcal{E})$, where $\mathcal{V} = \mathcal{R} \cup \mathcal{N} \cup \{v_0\}$ is the set of nodes representing \acp{RU}, cloud locations, and the \ac{CN}.
$\mathcal{E}$ denotes the set of transport links composing the \textit{crosshaul} network, interconnecting the \ac{CN}, cloud locations $\mathcal{N}$, and \acp{RU} $\mathcal{R}$.
These links support data transmission between \ac{RAN} functions hosted at different nodes, including communication between the \ac{CN} and cloud locations (\textit{backhaul}), between cloud locations hosting \acp{CU} and \acp{DU} (\textit{midhaul}), and between cloud locations and \acp{RU} (\textit{fronthaul})~\cite{morais:2023}.
Transport links are denoted by a tuple $(i,j) \in \mathcal{E}$ with a throughput capacity $e(i,j)$ representing the maximum data rate that can be transmitted over the transport link.
We denote by $\mathcal{P}^{m,n}$ the set of routing paths between nodes $m,n \in \mathcal{V}$. 
Each routing path $p \in \mathcal{P}^{m,n}$ is a sequence of transport links $(i, j) \in \mathcal{E}$ connecting nodes $m, n \in \mathcal{V}$ with latency $l(p)$, and can be used to transport \textit{backhaul}, \textit{midhaul}, or \textit{fronthaul} data, depending on the \ac{RAN} functions hosted by $m, n \in \mathcal{V}$.
We assume that the sets $\mathcal{N}$, $\mathcal{E}$, and $\mathcal{P}^{m,n}$, as well as the placement of \acp{CU} and \acp{DU} forming function chains with \acp{RU}, are known in the \textit{pre-failure} state.

\subsection{Failures in Disaggregated Mobile Networks}
\label{subsec:impact_of_failures}

Motivated by~\cite{almeida:24}, we assume the existence of a monitoring system based on heartbeats that identifies the set of compromised cloud locations after a failure.
This information, coupled with the \textit{pre-failure} placement of \ac{RAN} functions, identifies the \acp{RU} whose function chains have been interrupted, defining the set of disrupted \acp{RU}.
We denote the set of disrupted \acp{RU} as $\mathcal{R}_{dis}$, where $\mathcal{R}_{dis} \subseteq \mathcal{R}$, and the set of operational \acp{RU} as $\mathcal{R}_{op}$, where $\mathcal{R}_{op} = \mathcal{R} \setminus \mathcal{R}_{dis}$.
Similarly, the sets of operational cloud locations, transport links, and routing paths are denoted by $\mathcal{N}_{op}$, $\mathcal{E}_{op}$, and $\mathcal{P}_{op}^{m,n}$, respectively.
These sets characterize the network topology and resources during the \textit{in-failure} state.

We denote as $\mathcal{U}$ the set of users in the mobile network, where each user $u \in \mathcal{U}$ has a throughput demand $t(u)$.
When the service of \iac{RU} is disrupted, the users in its coverage area will attempt to re-associate with the network.
If users are within range of alternative operational \acp{RU}, they will re-associate with the network via those \acp{RU}.
Otherwise, the users remain disconnected due to the absence of an operational \ac{RU}.
To characterize user status in the \textit{in-failure} state, we define three categories:
\1 \textit{Non-affected Users}, whose serving \ac{RU} remains operational after a failure;
\2 \textit{Re-associated Users}, whose serving \ac{RU} was disrupted by a failure but who are re-associated to the network through an alternative operational \ac{RU}; and
\3 \textit{Disconnected Users}, whose serving \ac{RU} was disrupted by a failure and for whom no alternative operational \ac{RU} is available.
We denote the set of disconnected users as $\mathcal{U}_{dis}$, representing users that depend on resilience mechanisms to be re-associated with the network.

\begin{figure}[t]
  \centering
  \hfill
  \subfloat[a][\textit{Pre-failure} state.]{\includegraphics[width=0.235\textwidth]{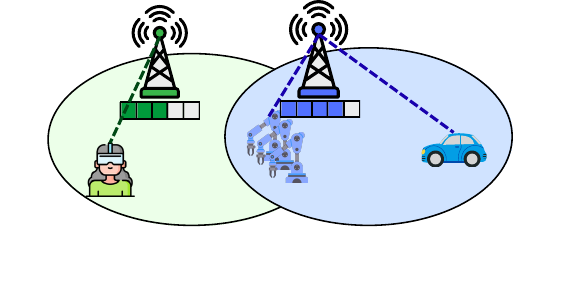}}\label{fig:pre_failure_association}
  \hfill
  \subfloat[b][\textit{In-failure} state.]{\includegraphics[width=0.245\textwidth]{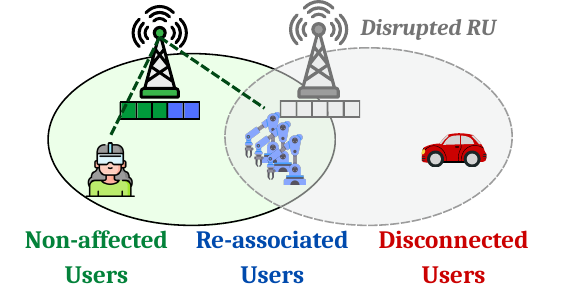}}\label{fig:in_failure_association}
  \hfill
  \caption{Impact of failures on users' connectivity. Non-affected users remain associated with their serving \ac{RU} from the \textit{pre-failure} state, re-associated users are served by alternative operational \acp{RU} during the \textit{in-failure} state, and disconnected users completely lose mobile network coverage.}
  \label{fig:types_of_users}
  \vspace{-1.3em}
\end{figure}

Figure~\ref{fig:types_of_users} illustrates the impact of failures on users' connectivity and network transmission resources.
While non-affected users remain associated with their serving \ac{RU} from the \textit{pre-failure} state, re-associated users are served by an alternative operational \acp{RU} with enough available resources during the \textit{in-failure} state.
However, due to the lack of coverage, disconnected users cannot be re-associated to the network during the \textit{in-failure} state, leaving the network unaware of their location and wireless channel conditions.

This partial loss of observability introduces uncertainty into the network state during failures.
In practice, some aspects of the network can still be deterministically observed, while others remain uncertain.
For instance, the set of compromised cloud locations and the resulting disrupted \acp{RU} can be identified through monitoring systems~\cite{Xing:2023,almeida:24}.
However, the locations and wireless channel conditions of disconnected users cannot be directly observed, as these users are no longer served by any \ac{RU}.
Consequently, although the availability of infrastructure resources is known, the location and wireless channel conditions of disconnected users remain uncertain during the recovery process.
Next, we characterize this uncertainty and formalize its role in the recovery process.

\subsection{Uncertainty in the Recovery Process}
\label{subsec:uncertainty}

We model the disconnected users' locations during the \textit{in-failure} state as random variables, whose realizations determine the resulting wireless channel conditions between disconnected users and disrupted \acp{RU}.
Based on~\cite{Chatterjee:2022}, let $\Psi_u$ denote the random location of a user $u \in \mathcal{U}_{dis}$.
The distance between user $u \in \mathcal{U}_{dis}$ and \ac{RU} $r \in \mathcal{R}$ is given by $\Delta_{u,r} = ||\Psi_u - \phi_r||_2$.

Let $\Xi_{u,r}$ denote the stochastic channel gain between user $u \in \mathcal{U}_{dis}$ and \ac{RU} $r \in \mathcal{R}$, and let $\Lambda(\Delta_{u,r})$ denote the path loss function calculated based on the distance between user and \ac{RU}.
The \ac{SINR} experienced by user $u \in \mathcal{U}_{dis}$ when served by \ac{RU} $r \in \mathcal{R}$ is given by
\begin{equation}
\label{eq:SINR}
    \Gamma_{u,r} =
    \frac{\Xi_{u,r} \ \Lambda(\Delta_{u,r}) \ p(r)}
    {\Upsilon_u + \sigma^2},
\end{equation}
\noindent where $\sigma^2$ is the variance of additive white Gaussian noise, and $\Upsilon_u$ is a random variable representing the cumulative downlink interference experienced by user $u \in \mathcal{U}_{dis}$ from other operating \acp{RU}, given by
\begin{equation}
\label{eq:interference}
    \Upsilon_u = 
    \sum_{q \in \mathcal{R}_{op} \setminus \{r\}}
    \Xi_{u,q} \ \Lambda(\Delta_{u,q}) \ p(q).
\end{equation}

Then, based on (\ref{eq:SINR}) and (\ref{eq:interference}), the expected spectral efficiency of user $u \in \mathcal{U}_{dis}$ served by \ac{RU} $r \in \mathcal{R}$ can be expressed as
\begin{align}
\label{eq:spec_eff}
    \Theta_{u, r}
    & = \log_2 \left( 1 + \Gamma_{u,r} \right) \nonumber \\
    & = \log_2 \left(
        1 +
        \frac{\Xi_{u,r} \, \Lambda(\Delta_{u,r}) \, p(r)}
        {I_u + \sigma^2}
    \right) \nonumber \\
    & = \log_2 \left(
        1 +
        \frac{\Xi_{u,r} \, \Lambda(\Delta_{u,r}) \, p(r)}
        {\sum\limits_{q \in \mathcal{R}_{op}\setminus\{r\}}
        \Xi_{u,q} \, \Lambda(\Delta_{u,q}) \, p(q) + \sigma^2}
    \right).
\end{align}

As discussed in~\cite{Rocha:2023}, the computational load of \ac{CU} and \ac{DU} instances, as well as the corresponding traffic demand over the transport network, scale with the bandwidth usage of their associated \acp{RU}.
However, the bandwidth utilization of disrupted \acp{RU} after recovery cannot be determined deterministically, as it depends on the uncertain locations and wireless channel conditions of disconnected users.
To account for this, we approximate the bandwidth usage of disrupted \acp{RU} after recovery based on two assumptions:
\1 disconnected users will be associated with the recovered \ac{RU} providing highest \ac{SINR};
and \2 bandwidth allocation to disconnected user will follow their throughput demand $t(u)$.

Let $P(u,r)$ denote the probability that a disrupted \ac{RU} $r \in \mathcal{R}_{dis}$, if recovered, provides the highest \ac{SINR} for user $u \in \mathcal{U}_{dis}$ among all disrupted \acp{RU}, defined as
\begin{equation}
\label{eq:association_probability}
P(u,r) = \mathbb{P}\!\left( \Gamma_{u,r} \ge \Gamma_{u,q} | \; \forall q \in \mathcal{R}_{dis} \setminus \{r\} \right).
\end{equation}
Moreover, let $T(u,r)$ denote the expected bandwidth required to serve user $u$ at \ac{RU} $r \in \mathcal{R}_{dis}$, after recovery, defined as
\begin{equation}
\label{eq:expected_user_bandwidth}
    T(u,r) = \frac{t(u)}{\mathbb{E}[\Theta_{u,r}]}.
\end{equation}
Then, based on Equations (\ref{eq:expected_user_bandwidth}) and (\ref{eq:association_probability}), the expected bandwidth utilization of a recovered \ac{RU} $r$ is given by
\begin{equation}
\label{eq:expected_ru_bandwidth}
    J(r) = 
    \sum_{u \in \mathcal{U}_{dis}} P(u,r) \, T(u,r).
\end{equation}

%% file: Sections/5-problem_formulation.tex
\section{Problem Formulation}
\label{sec:problem_formulation}

\begin{figure}[t]
  \centering
  \hfill
  \subfloat[a][\textit{In-failure} state.]{\includegraphics[width=0.15\textwidth]{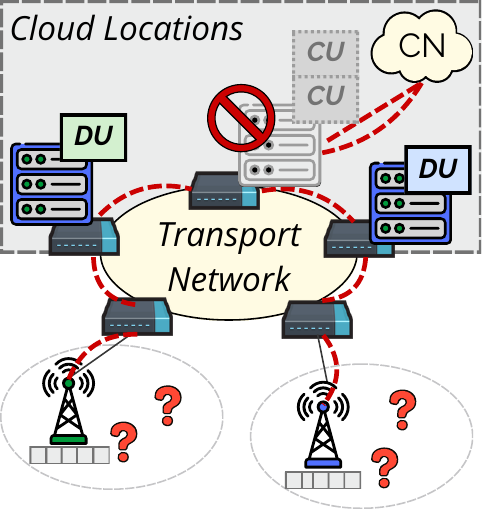}}\label{fig:in_failure_s}
  \hfill
  \subfloat[a][First stage.]{\includegraphics[width=0.15\textwidth]{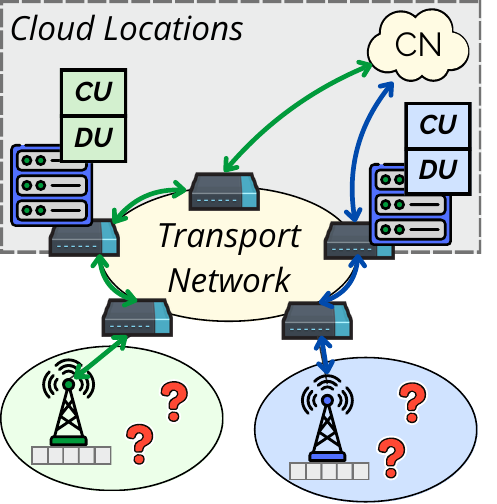}}\label{fig:first_stage}
  \hfill
  \subfloat[b][Second stage.]{\includegraphics[width=0.15\textwidth]{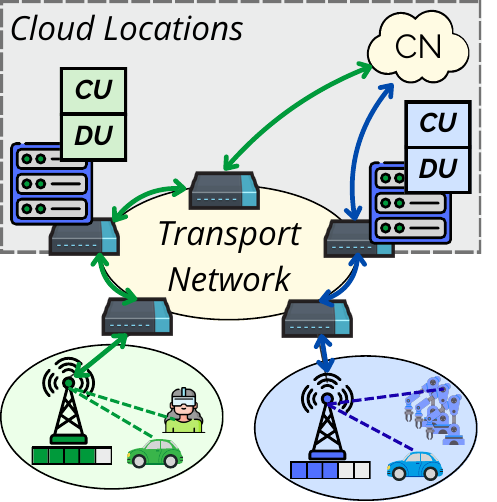}}\label{fig:second_stage}
  \hfill \vspace{+0.4em}
  \caption{Representation of our two-stage stochastic optimization framework. In the \textit{in-failure} state (a), both \acp{RU} are disrupted. In the first stage (b), under uncertainty about the locations and wireless channel conditions of disconnected users, both \acp{RU} are recovered. In the second stage (c), the uncertainty is resolved, and bandwidth is allocated to disconnected users.}
  \label{fig:two_stages}
  \vspace{-1em}
\end{figure}

Following the proposed system model in Section~\ref{sec:system_and_formulation}, our resilience mechanism must address system uncertainties in the \textit{in-failure} state, represented by random variables, which once resolved can lead to deterministic adaptations.
This sequential process leads to a two-stage stochastic optimization problem, where in the first stage, decisions are made under uncertainty, including:
\1 \emph{reinstantiation decisions}, which determine the reinstantiation of \ac{RAN} functions across cloud locations;
and \2 \emph{routing decisions}, which determine the paths used to transmit traffic between \ac{RAN} functions.
In the second stage, after uncertainty is resolved, \emph{bandwidth allocation decisions} determine how recovered \acp{RU} with restored function chains serve disconnected users.
In this context, the first stage corresponds to the \emph{here-and-now} problem, and the second stage corresponds to the \emph{recourse} problem~\cite{Birge:1997}.
The interdependence between the two stages characterizes a \ac{BTSP}, where first-stage decision variables are treated as fixed parameters in the second stage, and the second-stage objective is incorporated into the first-stage objective as an expected recourse value~\cite{manish:2025}. Figure~\ref{fig:two_stages} illustrates this two-stage stochastic optimization process. 

In the following, we present the mathematical formulation of the resilience problem leveraging the adaptive reinstantiation of \ac{RAN} functions in disaggregated mobile networks.

\subsection{Decision Variables}

To represent the reinstantiation decisions of the first stage, the decision variable\;$f_r^n\in\{0,1\}$\;indicates if the \ac{CU} associated with \ac{RU}\;$r\in\mathcal{R}_{dis}$\;is reinstantiated (or not) at an operational cloud location\;$n \in \mathcal{N}_{op}$, and the decision variable\;$g_r^n\in\{0,1\}$\;indicate if the \ac{DU} associated with \ac{RU}\;$r\in\mathcal{R}_{dis}$\;is reinstantiated (or not) at an operational cloud location\;$n\in\mathcal{N}_{op}$.
To represent the routing decisions regarding the communication between the reinstantiated \ac{RAN} functions, the decision variable\;$x_r^p \in \{0,1\}$ indicates if routing path\;$p \in \mathcal{P}_{op}^{v_0,n}$\;is selected (or not) to transport the \textit{backhaul} traffic of \ac{RU}\;$r \in \mathcal{R}_{dis}$.
Similarly, the decision variable\;$y_r^p \in \{0,1\}$ indicates if path\;$p \in \mathcal{P}_{op}^{m,n}$ is selected (or not) for the \textit{midhaul} traffic, and $z_r^p \in \{0,1\}$ indicate if path $p \in \mathcal{P}_{op}^{n,r}$ is selected (or not) for the \textit{fronthaul} traffic of \ac{RU}\;$r \in \mathcal{R}_{dis}$.
To represent the bandwidth allocation decisions of the second stage, the decision variable $b_r^u \in \mathbb{Z}_{+}$\;denotes the amount of bandwidth allocated in \ac{RU}\;$r \in \mathcal{R}_{dis}$\;to serve user\;$u \in \mathcal{U}_{dis}$\;after the restoration of its function chain.

\subsection{Objective Function}
\label{subsec:obj-function}

Following the definition of resilience in~\cite{WalidSaad:2024}, recovery performance can be expressed as the gap between the network utility in the \textit{pre-failure} state, $U(t_0)$, and the expected utility in the \textit{post-failure} state, $U(t_r)$ (see Fig.~\ref{fig:temporal_evolution_illustration}).
Thus, the objective of a generic resilience mechanism can be defined as
\begin{equation}
\label{eq:OF}
    \min \; U(t_0) - U(t_r),
\end{equation}
where the \textit{pre-failure} utility is considered an upper bound, i.e., $U(t_0) \geq U(t_r)$, and $U(t_0) = U(t_r)$ indicates full recovery.

In this work, we consider the \emph{aggregated throughput of all users as the network utility}, since it provides a measurable indication of the impact of failures in users' connectivity as shown in~\cite{Kaada:2022}.
Let $U({t_d})$ denote the aggregated throughput in the \textit{in-failure} state after stabilization.
We define the expected aggregated throughput in the \textit{post-failure} state as
\begin{equation}
    U(t_r) = U({t_d}) + \sum_{r \in \mathcal{R}_{dis}} \sum_{m,n \in \mathcal{N}_{op}} f_r^m \, g_r^n \, V(r),
\end{equation}
\noindent where $V(r)$ represents the expected aggregated throughput offered by \ac{RU} $r \in \mathcal{R}$, given by
\begin{equation}
\label{eq:expected_throughput}
    V(r) = \sum_{u \in \mathcal{U}_{dis}} b_r^u \, \mathbb{E} \Big [ \Theta_{u, r} \Big ].
\end{equation}

Based on (\ref{eq:OF}) -- (\ref{eq:expected_throughput}), the objective function of our problem formulation can be defined as follows:

\small
\begin{equation}
\label{eq:objective_function}
    \min_{f_r^m, g_r^n} \; U(t_0) - \bigg (U(t_d) + \sum_{r \in \mathcal{R}_{dis}} \sum_{m,n \in \mathcal{N}} f_r^m \, g_r^n \, \sum_{u \in \mathcal{U}_{dis}} b_r^u \mathbb{E} \Big [ \Theta_{u, r} \Big ] \bigg).
\end{equation}
\normalsize

The objective function defined in (\ref{eq:objective_function}) couples reinstantiation and bandwidth allocation decisions.
However, bandwidth allocation decisions depend on the realization of uncertainty regarding the locations and wireless channel conditions of disconnected users.
This dependence allows us to formulate the problem as a two-stage stochastic optimization problem, where:
\1 the first stage optimizes reinstantiation decisions ($f_r^m$ and $g_r^n$) to maximize recovery performance under uncertainty, determining which \acp{RU} are recovered;
and \2 the second stage optimizes bandwidth allocation decisions ($b_r^u$) to maximize the aggregate throughput of recovered \acp{RU} after uncertainty is realized.
We define the overall two-stage stochastic objective as follows:

\footnotesize
\begin{equation}
\label{eq:OF_BTSP}
    \min_{f_r^m, g_r^n}\!U(t_0)\!-\!\left (U(t_d)\!+\!\sum_{r \in \mathcal{R}_{dis}}\!\sum_{m,n \in \mathcal{N}_{op}}\!f_r^m g_r^n\!\max_{b_r^u}\sum_{u \in \mathcal{U}_{dis}}\!b_r^u \mathbb{E}\big[\Theta_{u,r}\big] \right ).
\end{equation}
\normalsize

\subsection{Problem Constraints}
\label{subsec:constraints}

Our proposed formulation is subject to two sets of constraints.
The \textit{first-stage constraints} define the feasibility of the \emph{here-and-now} problem, capturing the reinstantiation and routing requirements under uncertainty.
The \textit{second-stage constraints} define the feasibility of the \emph{recourse} problem, capturing the bandwidth allocation requirements after the realization of uncertainty.

\subsubsection{First-stage constraints}
\label{subsubsec:first-stage-constraints}
Let $C(r)$ and $D(r)$ denote the computational requirements of the \ac{CU} and \ac{DU} associated with \ac{RU} $r \in \mathcal{R}_{dis}$ under full bandwidth utilization~\cite{morais:2023}.
Accordingly, the following constraint ensures that the computational capacity of each operational cloud location is not exceeded, based on the expected bandwidth usage of disrupted \acp{RU} after recovery:
\begin{equation}
\label{eq:constraint1}
    \sum_{r \in \mathcal{R}_{dis}}
    \left(
        f_r^n \, C(r) + g_r^n \, D(r)
    \right)
    \frac{J(r)}{b(r)}
    \leq d(n),
    \quad \ \ \forall n \in \mathcal{N}_{op}.
\end{equation}

Let $T_p(i,j) \in \{0,1\}$, indicate if transport link $(i,j) \in \mathcal{E}_{op}$ belongs to routing path $p \in \mathcal{P}_{op}^{m,n}$.
Moreover, let $E(r,\mathrm{B})$, $E(r,\mathrm{M})$, and $E(r,\mathrm{F})$ denote the traffic load generated by \ac{RU} $r \in \mathcal{R}$ over the \textit{backhaul}, \textit{midhaul}, and \textit{fronthaul}, respectively, when operating at full bandwidth usage.
Based on the expected bandwidth utilization of recovered \acp{RU}, the following constraint ensures that the traffic carried over each transport link does not exceed its capacity:
\begin{align}
\label{eq:constraint2}
    &\sum_{r \in \mathcal{R}_{dis}} \sum_{m,n \in \mathcal{N}_{op}} f_r^m g_r^n \frac{J(r)}{b(r)} \Bigg ( \sum_{p \in \mathcal{P}^{v_0, m}_{op}} T_p(i,j) x_r^p E(r, B) + \nonumber \\
    & \sum_{p \in \mathcal{P}_{op}^{m,n}} T_p(i,j) y_r^p E(r, M) + \sum_{p \in \mathcal{P}_{op}^{n,r}} T_p(i,j) z_r^p E(r, F) \Bigg ) \nonumber \\
    & \qquad\qquad\qquad\qquad\qquad \ \ \leq e(i,j), \qquad \forall (i,j) \in \mathcal{E}_{op}.
\end{align}

To avoid unnecessary replication of \ac{RAN} functions during the restoration of interrupted function chains, each disrupted \ac{RU} $r \in \mathcal{R}_{dis}$ can be associated with at most one \ac{CU} instance and one \ac{DU} instance hosted at operational cloud locations:
\begin{align}
\label{eq:constraint3}
    \sum_{n \in \mathcal{N}_{op}} f_r^n \leq 1,
    &\qquad\qquad \forall r \in \mathcal{R}_{dis}, \\
    \sum_{n \in \mathcal{N}_{op}} g_r^n \leq 1,
    &\qquad\qquad \forall r \in \mathcal{R}_{dis}.
\end{align}

During the \textit{in-failure} state, a disrupted \ac{RU} may still have either its \ac{CU} or \ac{DU} (but not both) running at operational cloud locations, since the failure of a single \ac{RAN} function is sufficient to interrupt its function chain.
Let $K(r,n) \in \{0,1\}$ and $L(r,n) \in \{0,1\}$ indicate if cloud location $n \in \mathcal{N}_{op}$ hosts, respectively, the \ac{CU} and the \ac{DU} of \ac{RU} $r \in \mathcal{R}_{dis}$ during the \textit{in-failure} state.
To avoid redundant reinstantiation, existing \ac{CU} and \ac{DU} instances that remain operational must be preserved, preventing duplication of \ac{RAN} functions during recovery:
\begin{align}
\label{eq:constraint4}
    &f_r^n \geq K(r,n),
    \qquad\qquad \forall r \in \mathcal{R}_{dis}, \forall n \in \mathcal{N}_{op}, \\
    &g_r^n \geq L(r, n),
    \qquad\qquad \forall r \in \mathcal{R}_{dis}, \forall n \in \mathcal{N}_{op}.
\end{align}

If the functional chain of \ac{RU} $r \in \mathcal{R}_{dis}$ is restored, a routing path must be selected to transmit its \textit{backhaul}, \textit{midhaul}, and \textit{fronthaul} data.
The following constraints represent the relationships between reinstantiation and routing decisions:
\begin{align}
\label{eq:constraint5}
    &\sum_{p \in \mathcal{P}_{op}^{v_0,n}} x_r^p
    = \sum_{n \in \mathcal{N}_{op}} f_r^n,
    \qquad\qquad \forall r \in \mathcal{R}_{dis}, \\
    &\sum_{p \in \mathcal{P}_{op}^{n,m}} y_r^p
    = \sum_{m,n \in \mathcal{N}_{op}}\!f_r^m g_r^n,
    \qquad \forall r \in \mathcal{R}_{dis}, \\
    &\sum_{p \in \mathcal{P}_{op}^{m,r}} z_r^p
    = \sum_{m \in \mathcal{N}_{op}} g_r^m,
    \qquad\qquad \forall r \in \mathcal{R}_{dis}.
\end{align}

Let $Q(r,\mathrm{B})$, $Q(r,\mathrm{M})$, and $Q(r,\mathrm{F})$ denote the maximum latency threshold in the \textit{backhaul}, \textit{midhaul}, and \textit{fronthaul} for \ac{RU} $r \in \mathcal{R}_{dis}$, respectively \cite{morais:2023}.
Then, the selected routing paths to transport the data traffic between \ac{RAN} functions must satisfy these latency requirements:
\begin{align}
    &\sum_{p \in \mathcal{P}_{op}^{v_0,n}} x_r^p \, l(p) \leq Q(r,\mathrm{B}),
    \qquad\qquad \forall r \in \mathcal{R}_{dis}, \\
    &\sum_{p \in \mathcal{P}_{op}^{n,m}} y_r^p \, l(p) \leq Q(r,\mathrm{M}),
    \qquad\qquad \forall r \in \mathcal{R}_{dis}, \\
    &\sum_{p \in \mathcal{P}_{op}^{m,r}} z_r^p \, l(p) \leq Q(r,\mathrm{F}),
    \qquad\qquad  \forall r \in \mathcal{R}_{dis}. \label{eq:constraint6}
\end{align}

\subsubsection{Second-stage constraints}
\label{subsubsec:second-stage-constraints}

Given a feasible first-stage solution, the second stage determines the bandwidth allocation over recovered \acp{RU} to serve disconnected users.
In this stage, first-stage decision variables are treated as fixed parameters, and bandwidth allocation decisions are made after the realization of users' locations and wireless channel conditions.
A \ac{RU} is considered recovered if both its \ac{CU} and \ac{DU} are operational after the first-stage reinstantiation decisions.
Let $\mathcal{R}_{rec}$ denote the set of recovered \acp{RU}, i.e., the disrupted \acp{RU} whose function chains have been restored.
This\,set\,is\,defined\,as
\begin{equation}
\label{eq:constraint7}
    \mathcal{R}_{rec}
    =
    \left\{
    r \in \mathcal{R}_{dis} :
    \sum_{m,n \in \mathcal{N}_{op}} f_r^m g_r^n = 1
    \right\}. \nonumber
\end{equation}

The bandwidth allocated by recovered \acp{RU} must not exceed their capacity.
This requirement is enforced as follows:
\begin{equation}
\label{eq:constraint8}
    \sum_{u \in \mathcal{U}_{dis}} b_r^u \leq b(r),
    \qquad\qquad \forall r \in \mathcal{R}_{rec}.
\end{equation}

Given the reinstantiation decisions from the first stage, the computational load at each operational cloud location must not exceed its available capacity.
Since the computational requirements of \ac{CU} and \ac{DU} instances scale with the bandwidth allocated to their associated \acp{RU}~\cite{singh:21}, bandwidth allocation directly determines the resulting computational load.
Accordingly, the bandwidth allocation over recovered \acp{RU} must be constrained to ensure that the induced computational load remains within the capacity of each cloud location:
\begin{equation}
\label{eq:constraint9}
    \sum_{r \in \mathcal{R}_{rec}}
    \left(
        f_r^n \, C(r)\!+\!g_r^n \, D(r)
    \right)
    \frac{\sum\limits_{u\!\in\!\mathcal{U}_{dis}} b_r^u}{b(r)}
    \leq d(n),
 \forall n \in \mathcal{N}_{op}.
\end{equation}

Similarly, given the routing decisions from the first stage, the traffic carried over each operational transport link must not exceed its capacity.
Since the transmission load between \ac{RAN} functions scales with the bandwidth allocated to \acp{RU}~\cite{Rocha:2023}, bandwidth allocation directly determines the induced traffic on transport links.
Accordingly, bandwidth allocation over recovered \acp{RU} must be constrained to ensure that the resulting traffic load remains within the capacity of transport links:
\begin{align}
\label{eq:constraint10}
    &\sum_{r \in \mathcal{R}_{rec}} \sum_{m,n \in \mathcal{N}_{op}} f_r^m g_r^n \frac{\sum\limits_{u \in \mathcal{U}_{dis}} b_r^u}{b(r)} \Bigg ( \sum_{p \in \mathcal{P}^{v_0, m}_{op}} T_p(i,j) x_r^p E(r, B) + \nonumber \\
    & \sum_{p \in \mathcal{P}_{op}^{m,n}} T_p(i,j) y_r^p E(r, M) + \sum_{p \in \mathcal{P}_{op}^{n,r}} T_p(i,j) z_r^p E(r, F) \Bigg ) \nonumber \\
    & \qquad\qquad\qquad\qquad\qquad \ \ \leq e(i,j), \qquad \forall (i,j) \in \mathcal{E}_{op}.
\end{align}

The proposed formulation is a \ac{BTSP}~\cite{manish:2025}, where the first-stage objective function involves the expected value of the second-stage objective function.
The first stage determines reinstantiation and routing decisions under uncertainty, and the second stage allocates bandwidth to recovered \acp{RU} after the realization of users' locations and wireless channel conditions.
To solve this problem, we propose a \acf{SAA}-based solution that approximates the expected value using a finite set of sampled locations and wireless channel conditions for disconnected users.

%% file: Sections/6-solution.tex
\section{Sample Average Approximation Solution}
\label{sec:solution}

Our proposed \ac{SAA}-based solution consists of two stages. In the first stage, we determine the reinstantiation of \ac{RAN} functions to restore the function chains of disrupted \acp{RU} under uncertainty, using a finite set of scenarios $\Omega$, where each scenario $\omega \in \Omega$ represents a realization of disconnected users' locations and wireless channel conditions.
For each scenario $\omega \in \Omega$, the second stage is then solved independently to evaluate the corresponding recourse decisions.
Then, the two stages are solved iteratively.

Let $\psi_u^{(\omega)}$ denote the location of user $u \in \mathcal{U}_{dis}$ in scenario $\omega \in \Omega$.
For each pair of user $u \in \mathcal{U}_{dis}$ and \ac{RU} $r \in \mathcal{R}_{dis}$, the stochastic channel gain $\xi_{u,r}^{(\omega)}$ is independently sampled.
The distance between user $u \in \mathcal{U}_{dis}$ and \ac{RU} $r \in \mathcal{R}_{dis}$ in scenario $\omega \in \Omega$ is defined as $\delta_{u,r}^{(\omega)} = ||(\psi_u^{(\omega)} - \phi_r)||_2$.
Then, based on Eq.~(\ref{eq:SINR}), the \ac{SINR} of user $u \in \mathcal{U}$ to \ac{RU} $r \in \mathcal{R}$ in scenario $\omega \in \Omega$ is given by
\begin{equation}
    \gamma_{u,r}^{(\omega)} =
    \frac{\xi_{u,r}^{(\omega)} \Lambda\Big(\delta_{u,r}^{(\omega)}\Big) p(r)}
         {\upsilon_u^{(\omega)} + \sigma^2},
\end{equation}
where, based on Eq.~(\ref{eq:interference}), the downlink interference experienced by user is defined as
\begin{equation}
    \upsilon_u^{(\omega)} =
    \sum_{q \in \mathcal{R}_{op} \setminus \{r\}}
    \xi_{u,q}^{(\omega)} \Lambda\Big(\delta_{u,q}^{(\omega)}\Big) p(q),
\end{equation}
and, based on Eq.~(\ref{eq:spec_eff}) the spectral efficiency is:
\begin{equation}
    \theta_{u, r}^{(\omega)} = \log_2\!\left(1 + \gamma_{u,r}^{(\omega)}\right).
\end{equation}

Then, based on Eq.~(\ref{eq:expected_user_bandwidth}), the bandwidth requirement to user $u \in \mathcal{U}$ at \ac{RU} $r \in \mathcal{R}$ on scenario $\omega \in \Omega$ is:
\begin{equation}
    T_{\omega}(u, r) = \frac{t(u)}{\theta_{u,r}^{(\omega)}}.
\end{equation}

Since the \ac{SINR} values depend on the uncertain locations and wireless channel conditions of disconnected users, the resulting constraints cannot be expressed in closed form.
To address this, we construct an empirical estimator based on \ac{SINR} realizations to parameterize the constraints of the first stage~\cite{Carlinet:2019}.
Based on Eq.~(\ref{eq:association_probability}), we assume that user $u \in \mathcal{U}_{dis}$ will be associated with the recovered \ac{RU} with the highest \ac{SINR}.
We define this as
\begin{equation}
    P_\omega(u,r) = \mathbbm{1}_{\left\{\gamma_{u,r}^{(\omega)} \ge \gamma_{u,q}^{(\omega)}, \; \forall q \in \mathcal{R}_{dis} \setminus \{r\} \right\}},
\end{equation}
where $\mathbbm{1}_{\{\cdot\}}$ denotes the indicator function, which equals one if the condition inside the brackets is satisfied or zero otherwise.

Based on Eq.~(\ref{eq:expected_ru_bandwidth}), the expected bandwidth utilization of \ac{RU} $r \in \mathcal{R}_{dis}$ in a scenario $\omega \in \Omega$, if recovered, is defined as
\begin{equation}
J_{\omega}(r) = \sum_{u \in \mathcal{U}_{dis}} P_{\omega}(u, r) \ T_{\omega}(u, r).
\end{equation}

By approximating the expectation over uncertainty with a finite set of sampled realizations (scenarios), the original stochastic problem is represented as \iac{SAA}-based solution, where the first stage can be formulated as follows:

\small
\begin{align}
\label{eq:saa_objective1}
&\minimize_{f_r^m, g_r^n}
\Bigg[
U(t_0)
- U(t_d) -
\frac{1}{|\Omega|}
\sum_{\omega \in \Omega}
\Bigg(
\sum_{r\in\mathcal{R}_{dis}}\!
\sum_{m,n\in\mathcal{N}_{op}}\!
f_r^m\!g_r^n\! \nonumber \\
& \qquad\qquad\qquad\quad\quad\quad\quad\quad\quad\quad\quad\quad\quad
\sum_{u \in \mathcal{U}_{dis}}\!
J_{\omega}(r) \theta_{u,r}^{(\omega)}
\Bigg)
\Bigg]
\end{align}

\noindent subject to:
\begin{align}
& \sum_{r \in \mathcal{R}_{rec}} \left(
        f_r^n \, C(r)\!+\!g_r^n \, D(r)
    \right)
    \frac{\frac{1}{|\Omega|}\sum\limits_{\omega \in \Omega} J_\omega(r)}{b(r)}
    \leq d(n),
 \forall n \in \mathcal{N}_{op}, \label{eq:saa_c1} \\
 &\sum_{r \in \mathcal{R}_{dis}} \sum_{m,n \in \mathcal{N}_{op}} f_r^m g_r^n \frac{\frac{1}{|\Omega|}\sum\limits_{\omega \in \Omega} J_\omega(r)}{b(r)} \Bigg ( \sum_{p \in \mathcal{P}^{v_0, m}_{op}} T_p(i,j) x_r^p E(r, B) + \nonumber \\
    &\qquad\qquad \sum_{p \in \mathcal{P}_{op}^{m,n}} y_r^p T_p(i,j) E(r, M) + \sum_{p \in \mathcal{P}_{op}^{n,r}} z_r^p T_p(i,j) E(r, F) \Bigg ) \nonumber \\
    & \qquad\qquad\qquad\qquad\qquad\qquad\quad \leq e(i,j), \qquad \forall (i,j) \in \mathcal{E}_{op}, \label{eq:saa_c2} \\
  &\text{ and Equations (\ref{eq:constraint3}) -- (\ref{eq:constraint6}).} \nonumber
\end{align}
\normalsize

Equation~(\ref{eq:saa_objective1}) defines the objective function of our first stage \ac{SAA}-based solution.
While~(\ref{eq:saa_c1}) is the computational capacity constraint of cloud locations, analogous to~(\ref{eq:constraint1}), and~(\ref{eq:saa_c2}) is the capacity constraint of transport links, analogous to~(\ref{eq:constraint2}).

Note that the first stage of our \ac{SAA}-based solution does not account for bandwidth allocation decisions, since the locations and wireless channel conditions of disconnected users are uncertain.
Thus, to solve the bandwidth allocation, we formulate the second stage of our \ac{SAA}-based solution as:

\small
{
\begin{align}
\label{eq:saa_objective2}
\maximize_{b_r^{u, \omega}} \sum_{\omega \in \Omega} \sum_{r \in \mathcal{R}_{rec}} \sum_{u \in \mathcal{U}_{dis}}  b_r^{u,(\omega)} \theta_{u,r}^{(\omega)}.
\end{align}
\noindent Subject to:
\begin{align}
    &\sum_{u \in \mathcal{U}_{dis}} b_r^{u,(\omega)} \leq b(r),
    \qquad\qquad \forall r \in \mathcal{R}_{rec}, \omega \in \Omega, \label{eq:saa_c3}\\
    &\sum_{r \in \mathcal{R}_{rec}}
    \left(
        f_r^n \, C(r)\!+\!g_r^n \, D(r)
    \right)
    \frac{\sum\limits_{u\!\in\!\mathcal{U}_{dis}} b_r^{u,(\omega)}}{b(r)}
    \leq d(n), \forall n \in \mathcal{N}_{op}, \omega \in \Omega. \label{eq:saa_c4} \\
    &\sum_{r \in \mathcal{R}_{rec}} \sum_{m,n \in \mathcal{N}_{op}} f_r^m g_r^n \frac{\sum\limits_{u \in \mathcal{U}_{dis}} b_r^{u,(\omega)}}{b(r)} \Bigg ( \sum_{p \in \mathcal{P}^{v_0, m}_{op}} T_p(i,j) x_r^p E(r, B) + \nonumber \\
    & \sum_{p \in \mathcal{P}_{op}^{m,n}} y_r^p T_p(i,j) E(r, M) + \sum_{p \in \mathcal{P}_{op}^{n,r}} z_r^p T_p(i,j) E(r, F) \Bigg ) \nonumber \\
    & \qquad\qquad\qquad\qquad\qquad \leq e(i,j), \qquad \forall (i,j) \in \mathcal{E}_{op}, \omega \in \Omega. \label{eq:saa_c5}
\end{align}
}
\normalsize

Equation~(\ref{eq:saa_objective2}) defines the objective function of our second stage \ac{SAA}-based solution.
While (\ref{eq:saa_c3}) is the bandwidth capacity constraints of recovered \acp{RU}, analogous to~(\ref{eq:constraint7}). 
Equation~(\ref{eq:saa_c4}) is the computational capacity constraint of cloud locations, analogous to~(\ref{eq:constraint9}); and~(\ref{eq:saa_c5}) is the operational transport links capacity constraint, analogous to~(\ref{eq:constraint10}).

Our \ac{SAA}-based solution yields a deterministic equivalent problem that can be solved with a deterministic solver, e.g., CPLEX.
The two stages are solved iteratively, with first-stage decisions as fixed parameters in the second-stage problem. This structure allows parallel execution of the second stage, significantly improving its computational efficiency. We describe our \ac{SAA}-based solution in Algorithm~\ref{alg:saa}.

\begin{algorithm}[t]
\small
\caption{SAA-based solution}
\label{alg:saa}
\SetKwInOut{Input}{Input}
\SetKwInOut{Output}{Output}

\Input{$\mathcal{R}_{dis}$, $\mathcal{U}_{dis}$, $\mathcal{N}_{op}$, $\mathcal{E}_{op}$, $\mathcal{P}_{op}^{m,n}$, and $\Omega$}
\Output{Reinstantiation decisions $(f_r^m, g_r^n)$ and bandwidth allocations decisions $b_r^{u,(\omega)}$}

Generate $|\Omega|$ scenarios of user locations and channel conditions\;

\For{$\omega \in \Omega$}{
    Compute $\theta_{u,r}^{(\omega)}$, $T_{\omega}(u,r)$, $P_{\omega}(u,r)$, and $J_{\omega}(r)$\;
}

Solve the first-stage \ac{SAA} problem to obtain $\mathcal{R}_{rec}$\;

\For(\textbf{in parallel}){$\omega \in \Omega$}{
    Set fixed values for $(f_r^m, g_r^n)$ and $\mathcal{R}_{rec}$\;
    Solve the second-stage problem for $\omega \in \Omega$ to obtain bandwidth allocation $b_r^{u,(\omega)}$\;
}

Aggregate results across scenarios\;

\Return $(f_r^m, g_r^n)$ and $\{b_r^{u,(\omega)}\}_{\omega \in \Omega}$\;
\end{algorithm}

%% file: Sections/7-evaluation.tex
\section{Evaluation}
\label{sec:evaluation}

In this section, we evaluate the performance of our resilience mechanism, reinstantiating \ac{RAN} functions to restore the function chain of disrupted \acp{RU} and recover service for disconnected users.
First, we describe our simulation setup and failure scenarios.
Then, we compare our solution against existing approaches in terms of recovery performance, throughput resilience, temporal recovery evolution, and resource utilization, under different failure severities and traffic demands. 

\subsection{Simulation Environment}
\label{subsec:simulation_environment}

\begin{figure}[t]
    \centering
    \includegraphics[width=0.80\columnwidth]{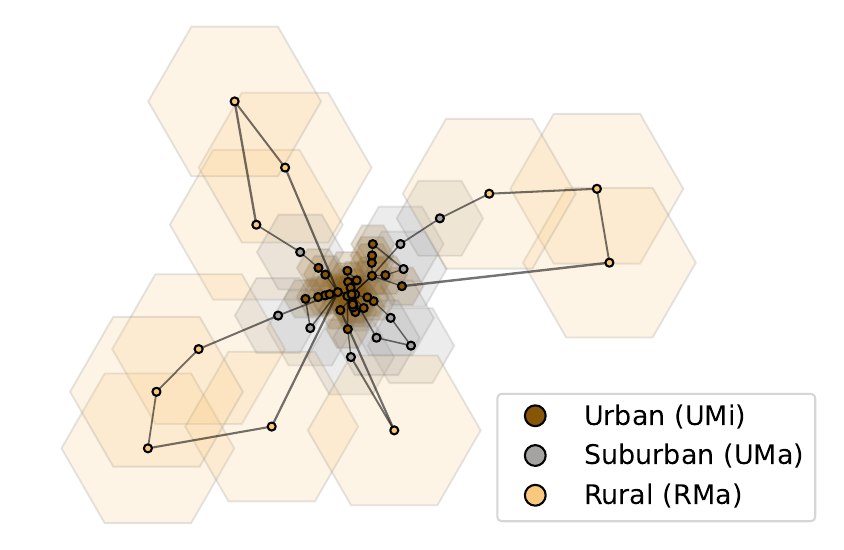}
    \caption{Real mobile network topology comprising 50 \acp{RU} distributed across urban, suburban, and rural regions, interconnected through a ring-based \textit{crosshaul} transport network~\cite{morais:2023}.}
    \label{fig:topology}
\end{figure}

We developed a simulator to evaluate our solution, leveraging a real-world network topology of a city in Italy~\cite{5Gxcrosshaul:17}, illustrated in Fig.~\ref{fig:topology}.
The topology comprises 50 \acp{RU}, distributed across urban, suburban, and rural regions, each co-located with a cloud location and interconnected through a ring-based transport network~\cite{morais:2023}.
To capture the different propagation characteristics of each region, we adopted the \ac{UMi}, \ac{UMa}, and \ac{RMa} channel models for \acp{RU} deployed in urban, suburban, and rural areas, respectively~\cite{sun:2016,maccartney:2017}.
In addition, we summarize the different cell bandwidth, transmit power, cell radius, user density, and path-loss exponent of each region
in Table~\ref{tab:evaluation}.
Moreover, to establish a feasible \textit{pre-failure} network deployment as the baseline for evaluating the recovery process, we use the simulator presented in~\cite{morais:2023} to determine the placement of \ac{RAN} functions and the selection of functional splits for each \ac{RU}, considering 3GPP Split 2 (CU--DU interface)~\cite{TS38401:2025} and O-RAN Split 7.2x (DU--RU interface)~\cite{openRAN:22}.

\begin{table}[t]
\scriptsize
\centering
\begin{tabular}{|l|ccccc|}
\hline
Region & Bandwidth & \begin{tabular}[c]{@{}c@{}}Transmit\\ power\end{tabular} & \begin{tabular}[c]{@{}c@{}}Cell\\ radius\end{tabular} & \begin{tabular}[c]{@{}c@{}}User\\ density\end{tabular} & \begin{tabular}[c]{@{}c@{}}Path-loss\\ exponent\end{tabular} \\ \hline
Urban (UMi)     & 100 MHz    & 30 dBm     & 250 m   & 200/km$^2$    & 2.0 \\
Suburban (UMa)  & 80 MHz     & 40 dBm     & 500 m   & 20/km$^2$    & 2.8 \\
Rural (RMa)     & 40 MHz     & 46 dBm     & 1000 m  & 6/km$^2$   & 2.31 \\ \hline
\end{tabular}
\caption{Characteristics of \acp{RU} in different regions.}
\label{tab:evaluation}
\vspace{-2em}
\end{table}

We consider users to be spatially distributed across the network coverage area according to a \acf{HPPP}, with region-specific densities presented in Table~\ref{tab:evaluation}, based on~\cite{ETSI203228}.
This stochastic spatial distribution models the uncertainty in users' locations after failures, as the network cannot observe the positions of disconnected users when \acp{RU} with interrupted function chains are unable to collect channel-state information.

\begin{figure}[t]
  \centering
  \hfill
  \subfloat[a][Disrupted RUs.]{\includegraphics[width=0.48\columnwidth]{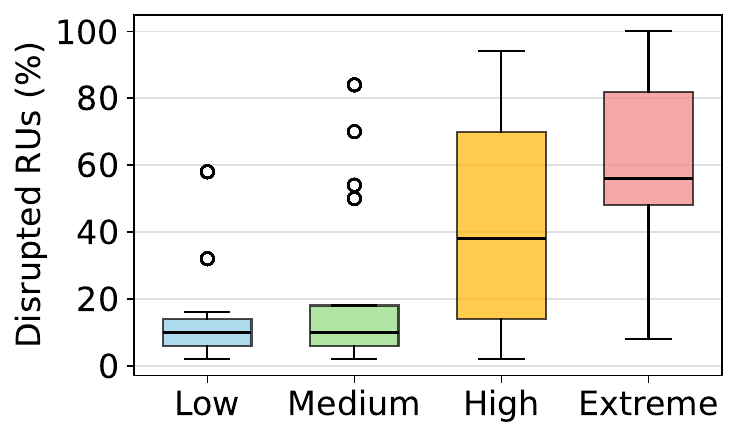}\label{fig:disrupted_RUs}}
  \hfill
  \subfloat[b][Disconnected Users.]{\includegraphics[width=0.48\columnwidth]{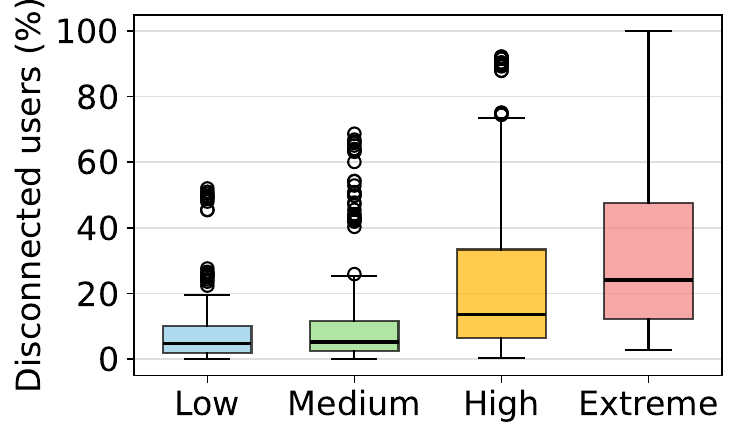}\label{fig:disconnected_users}}
  \hfill \vspace{-0.5em}
  \caption{Impact of the failure scenarios with different severity levels 
  in terms of disrupted \acp{RU} and disconnected users.}
  \label{fig:failure_impact}
  \vspace{-1em}
\end{figure}

Following~\cite{almeida:2025}, we evaluate the recovery performance of our resilience mechanism under four failure severity levels, in which different percentages of random cloud locations are compromised: 
low severity (5\%), medium severity (10\%), high severity (25\%), and extreme severity (50\%).
These failure scenarios enable us to examine the performance of our resilience mechanism under various conditions.
We consider 10 instances for each severity level, each corresponding to a different realization of compromised cloud locations.
For each instance, we generate 30 realizations of user spatial distribution following the \ac{HPPP}, forming the set of scenarios $\Omega$ used in our \ac{SAA}-based solution.
This experimental design results in 1,200 problem instances. Figure~\ref{fig:failure_impact} illustrates the impact of each severity level on the number of disrupted \acp{RU} and disconnected users, which increase with failure severity, as the number of interrupted function chains grows.
Moreover, at high and extreme severity levels, the variability in the number of disrupted \acp{RU} and disconnected users increases significantly due to cascading failures.

We compare the performance of our resilience mechanism against three benchmarks in the literature:
\1 \textit{Kaada}~\cite{KAADA:24}, which adapts the transmit power and antenna tilt of operational \acp{RU} after failures to expand their coverage area to recover disconnected users; 
\2 \textit{\ac{DFR}}~\cite{almeida:2025}, corresponding to our previous work, which reinstantiates \ac{RAN} functions without accounting for uncertainty in wireless channel conditions and locations of users; and 
\3 the \textit{\ac{WS}} solution, which provides an optimal theoretical upper bound by assuming perfect knowledge of disconnected users' locations and wireless channel conditions.

\subsection{Recovery Performance}

First, we assess the recovery performance of deterministic and stochastic resilience mechanisms by evaluating their \textit{recovered throughput}, defined as the increase in aggregate network throughput from the \textit{in-failure} state (after failure) to the \textit{post-failure} state (after recovery).

\begin{figure}
    \centering
    \includegraphics[width=1\linewidth]{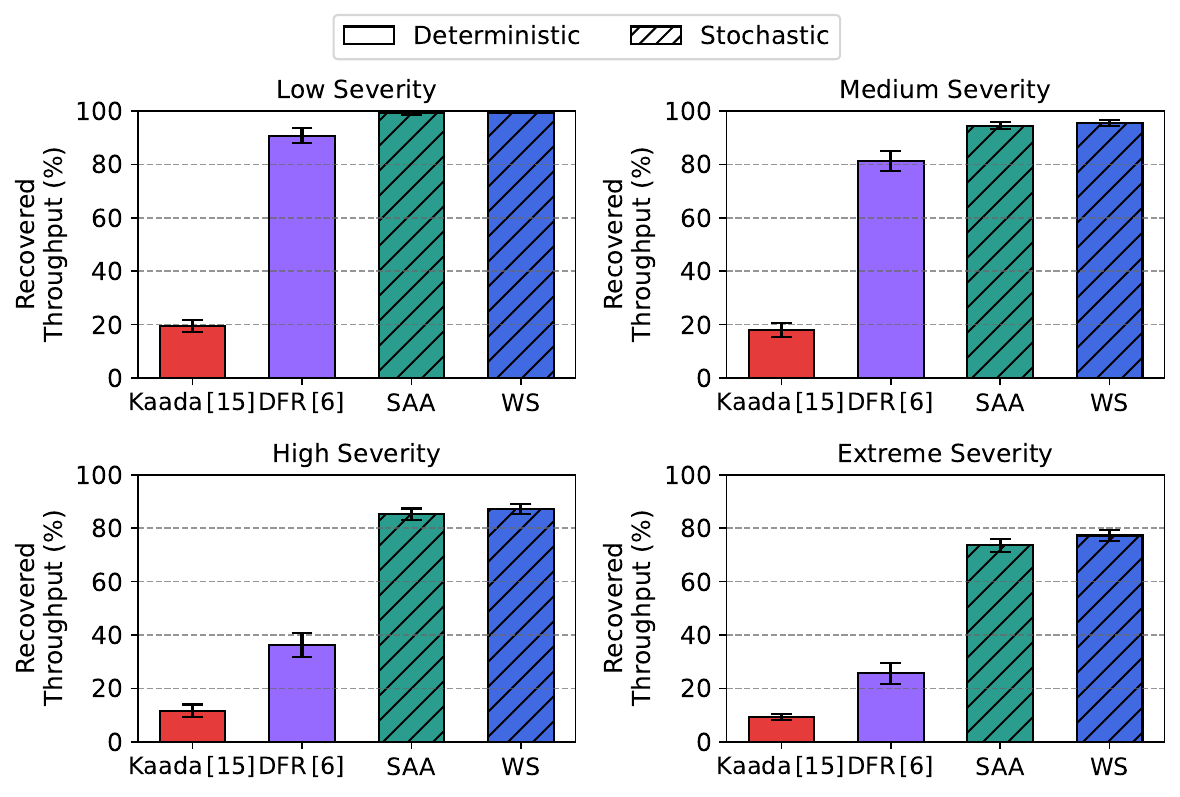}
    \caption{Recovery performance of deterministic and stochastic resilience mechanisms under different severity levels.
    Across all scenarios, our \ac{SAA} solution achieves 
    recovered throughput higher than the baseline approaches and close to optimal. 
    }
    \label{fig:utility_comparison_bars}
    \vspace{-1em}
\end{figure}

Figure~\ref{fig:utility_comparison_bars} shows the recovered throughput of each resilience mechanism under low, medium, high, and extreme severity levels. 
\textit{Kaada}~\cite{KAADA:24} achieves the lowest performance across all scenarios, presenting 10--20\% of recovered throughput since it relies solely on the remaining operational \acp{RU} to restore connectivity for disconnected users.
The \ac{DFR}~\cite{almeida:2025} achieves 80\% and 93\% of recovered throughput under low and medium severity levels, respectively, but its performance decreases by 2--4 times at high and extreme severity levels due to the larger number of disrupted \acp{RU} and disconnected users (see Fig.~\ref{fig:failure_impact}).
Our proposed solution achieves the highest recovered throughput across all scenarios, with gains of 64--80\% over \textit{Kaada} and 9--48\% over \ac{DFR}, while remaining within 0.3--3.7\% of the theoretical upper bound, \ac{WS}.
The performance gain of our proposed solution reflects the benefits of considering statistical information about the \textit{post-failure} state rather than \textit{pre-failure} information during recovery, demonstrating that resilience mechanisms that account for uncertainty in users' locations and wireless channel conditions enable better recovery, particularly under severe failure conditions.

\subsection{Responding to Failures and Restoring Service}

\begin{figure}
    \centering
    \includegraphics[width=1\linewidth]{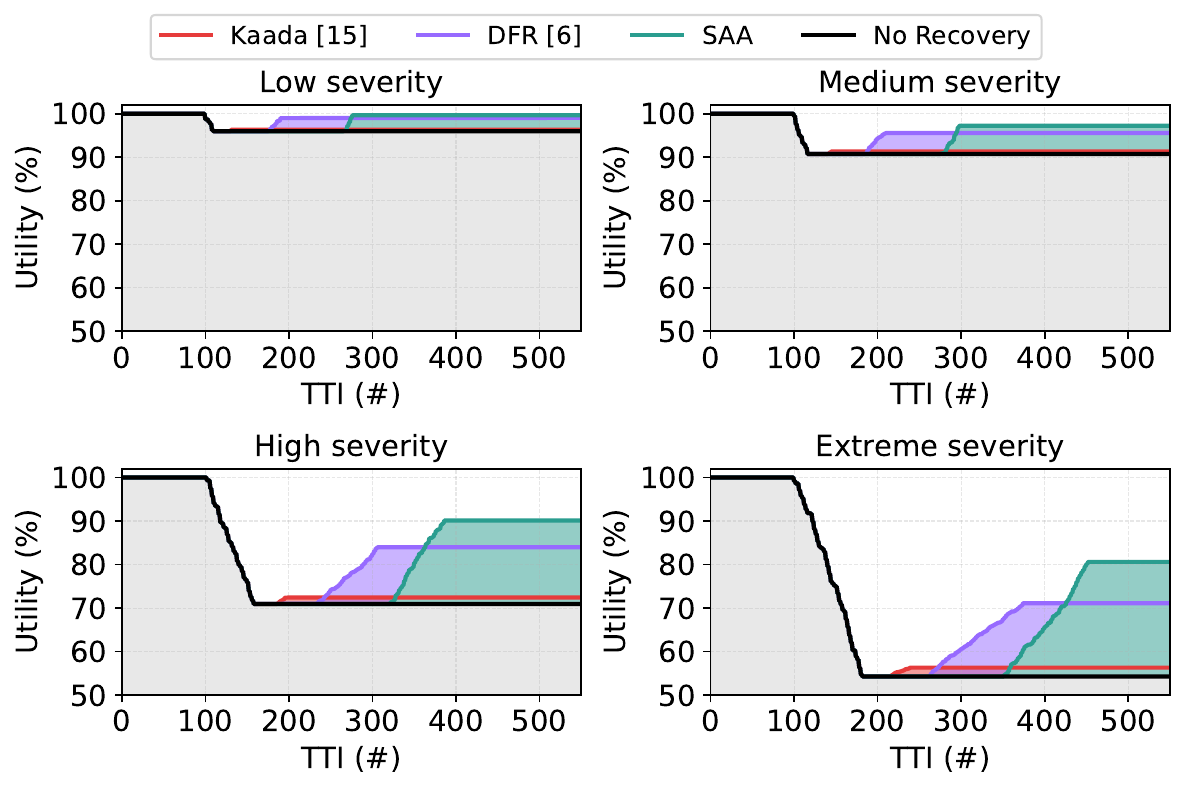}
    \caption{Comparison of the temporal evolution of network utility across all resilience mechanisms. After the disruption, the network utility stabilizes, and the mobile operator triggers the resilience mechanisms that recover it.}
    \label{fig:temporal_evolution}
\end{figure}

Figure~\ref{fig:temporal_evolution} shows the evolution of network utility over time across different failure severity levels, evaluated per \ac{TTI} with \unit[15]{KHz} subcarrier spacing~\cite{TS38211:2025}.
This presentation allows us to observe the dynamics of utility recovery over time and evaluate trade-offs between performance and recovery time.
As we can observe, network utility degrades as failure severity increases, reflecting the impact of compromised cloud locations.
\textit{Kaada} achieves faster recovery due to its lower complexity, followed by \ac{DFR}, while our \ac{SAA}-based solution incurs higher computational cost but achieves the highest utility recovery, with gains of up to 25\%.
These results reveal a trade-off between recovery time and recovery performance: simpler mechanisms enable faster reactions, whereas more sophisticated approaches achieve higher post-recovery utility.
This suggests that future resilience mechanisms could jointly consider both dimensions, potentially through adaptive strategies or hybrid short- and long-term recovery actions~\cite{WalidSaad:2024}.
Without loss of generality, we assumed a fixed wait period of \unit[40]{ms} before triggering recovery, as per~\cite{Xing:2023}, while the investigation of failure detection mechanisms remains part of future work.

\subsection{Throughput Resilience and Computing Resource Usage}

\begin{figure}[t]
  \centering
  \hfill
  \subfloat[a][Throughput resilience.]{\includegraphics[width=0.24\textwidth]{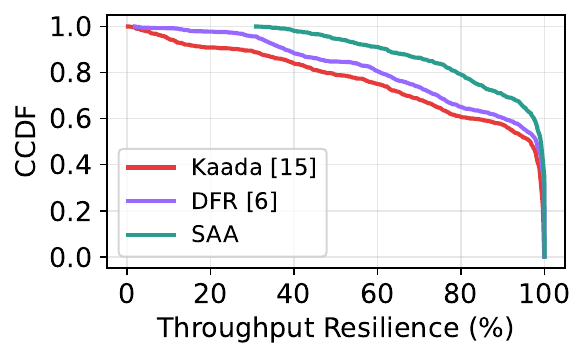}\label{fig:CCDF_conservative}}
  \hfill
  \subfloat[b][CPU usage.]{\includegraphics[width=0.24\textwidth]{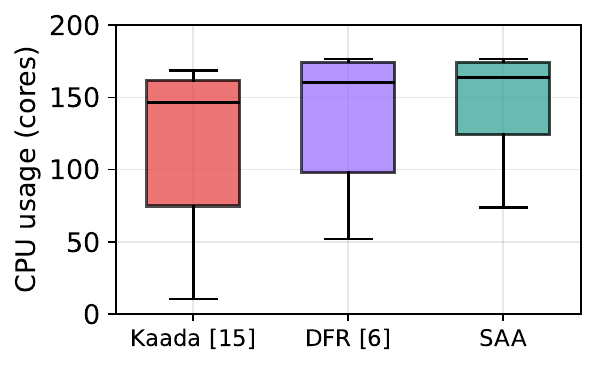}\label{fig:CPU_usage}}
  \hfill
  \vspace{-0.5em}
  \caption{Comparison of the distributions of throughput resilience and processing resource utilization of the different resilience mechanisms across all problem instances.}
  \label{fig:throughput_resilience}
  \vspace{-1em}
\end{figure}

We evaluate how each resilience mechanism restores network performance relative to the \textit{pre-failure} state, while analyzing the processing resources required to achieve this recovery.
We adopt the \textit{throughput resilience} metric~\cite{almeida:2025}, defined as the ratio between the aggregate network throughput in the \textit{post-failure} state (after recovery) and that in the \textit{pre-failure} state (before failure).

Figure~\ref{fig:throughput_resilience} presents the \ac{CCDF} of throughput resilience, together with the corresponding processing resource usage across cloud locations after recovery for each resilience mechanism.
\textit{Kaada}~\cite{KAADA:24} achieves the lowest throughput resilience, guaranteeing at least $20\%$ throughput resilience for $90\%$ of the problem instances, while maintaining the lowest processing usage, as it relies solely on the remaining operational \acp{RU}.
In contrast, \ac{DFR}~\cite{almeida:2025} guarantees $40\%$ throughput resilience for $90\%$ of the instances, at the cost of increased processing resource usage, which may lead to higher operational costs during recovery.
Our \ac{SAA}-based solution achieves the highest performance, guaranteeing $70\%$ throughput resilience for $90\%$ of the instances, while presenting only $11\%$ of increase in CPU usage, as it restores a larger number of \acp{RU}, requiring the deployment of additional \ac{CU} and \ac{DU} instances across cloud locations.
These results highlight the trade-off between recovery performance and computational cost, where the higher performance of our \ac{SAA}-based solution stems from its ability to restore the service for a higher number of \acp{RU}, which requires deploying more \ac{RAN} function instances and, consequently, increases processing resource consumption.

\subsection{Recovery Performance Under Different User Demand}

\begin{figure}[t]
  \centering
  \vfill
  \subfloat[a][Low demand (250 active users).]{\includegraphics[width=0.98\columnwidth]{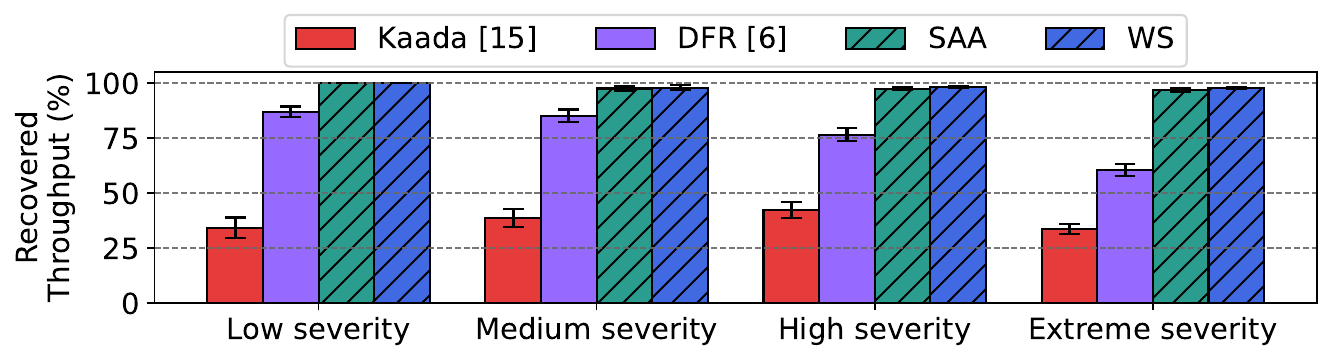}\label{fig:cons_adap_low}}
  \vfill
  \subfloat[b][Medium demand (500 active users).]{\includegraphics[width=0.98\columnwidth]{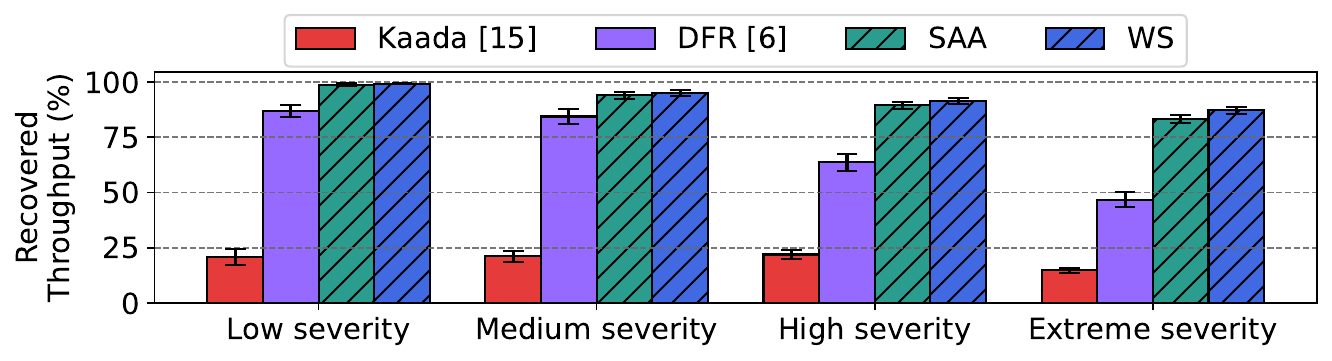}}\label{fig:cons_adap_mid}
  \vfill
  \subfloat[c][High demand (1000 active users).]{\includegraphics[width=0.98\columnwidth]{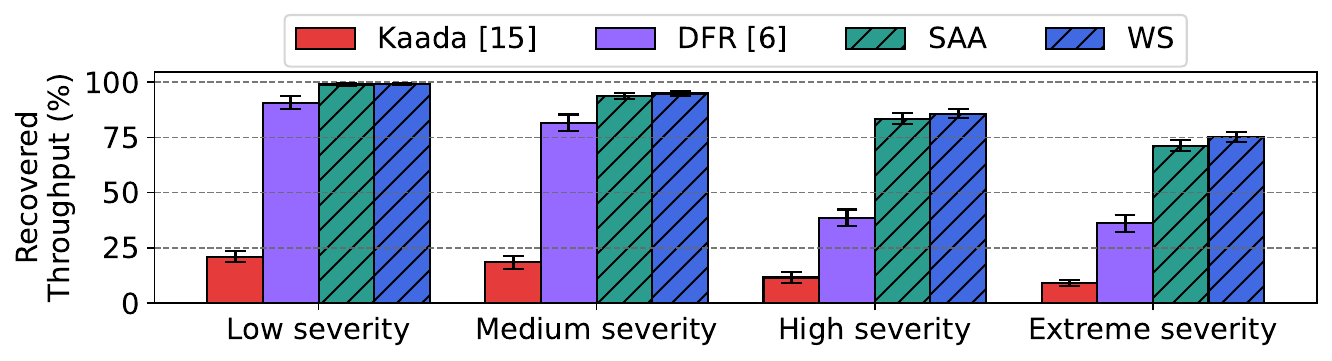}}\label{fig:cons_adap_high}
  \vfill
  \caption{Recovery performance of different resilience mechanisms under conservative and adaptive policies.}
  \label{fig:recovered_all_demand}
  \vspace{-1em}
\end{figure}

The traffic load in a mobile network varies with the number of active users, their throughput requirement, and mobility patterns~\cite{Depeng:17}, impacting the \ac{RU} bandwidth usage and the computational and transport requirements of their function chains, as captured in Eqs.~(\ref{eq:constraint1}) and~(\ref{eq:constraint2}), which can affect both the recovery performance and feasibility of solutions. 
To understand the impact of the traffic load, we compared the aggregate recovered throughput achieved by the different resilience mechanisms under three demand scenarios based on the number of active users: low (250 users), medium (500 users), and high (1000 users).

Figure~\ref{fig:recovered_all_demand} shows the recovered throughput achieved by each resilience mechanism under different demand scenarios.
Our \ac{SAA}-based solution consistently outperforms \textit{Kaada} across all scenarios, achieving performance gains of 55--65\% under low demand and 62--77\% under medium and high demand.
Compared to \ac{DFR}, our solution achieves gains of 10--36\% under low and medium demand, and up to 45\% under high demand, while remaining within 0.5--1.8\% of the theoretical upper bound, \ac{WS}.
These results show that accounting for uncertainty in user locations and wireless channel conditions enables more adaptive recovery decisions, leading to improved performance, particularly under high-demand scenarios where resource constraints are more pronounced.

\subsection{Impact of \acp{RU} Recovery Across Regions}

\begin{figure}
    \centering
    \includegraphics[width=0.88\linewidth]{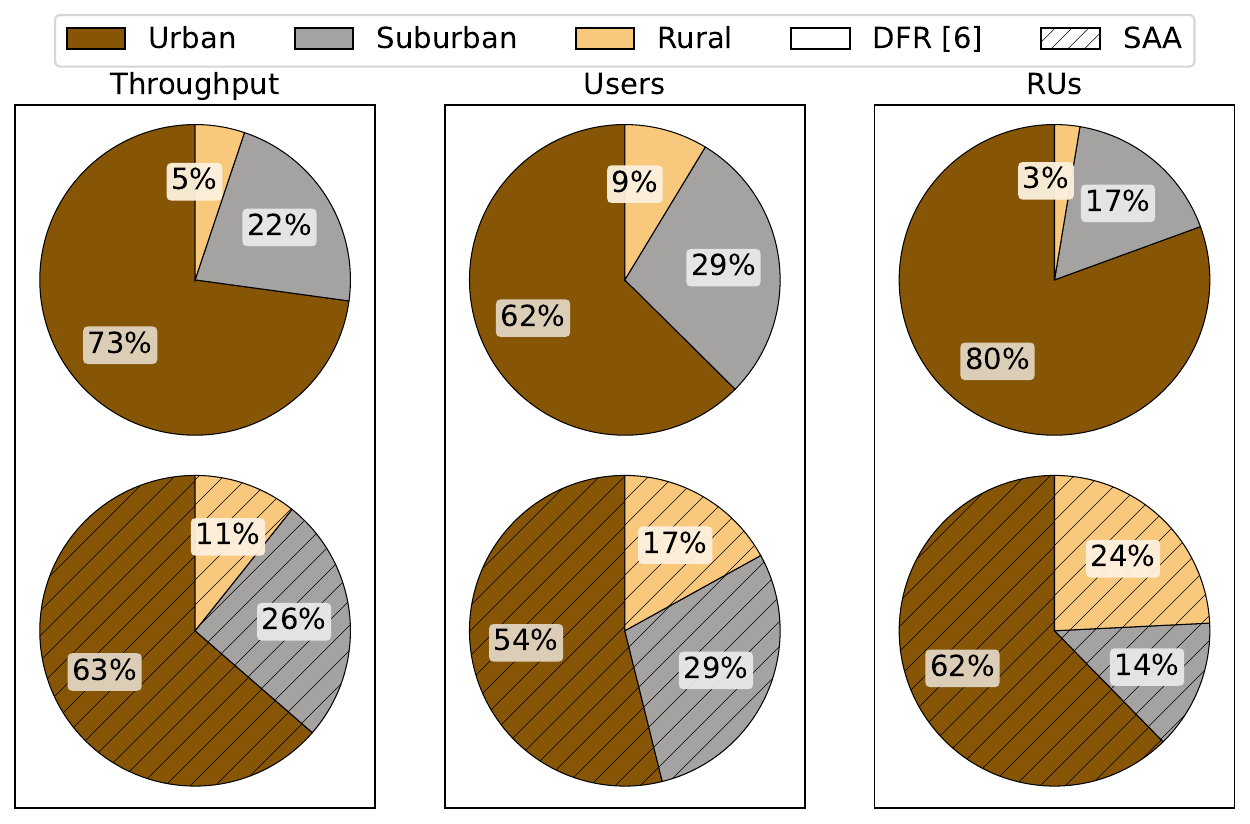}
    \caption{Comparison of recovery performance across different region types. The deterministic approach prioritizes urban 
    regions, while our stochastic solution achieves higher throughput resilience and with a more balanced recovery across  regions.}
    \label{fig:different_utilities}
\end{figure}

Finally, we evaluate how the recovered throughput achieved by deterministic and stochastic resilience mechanism translates into recovering \acp{RU} and reconnecting users across different region types.
This provides insight into how resilience mechanisms prioritize \ac{RU} recovery and user service restoration under limited network resources.
Figure~\ref{fig:different_utilities} presents the distribution of recovered throughput, users, and \acp{RU} across urban, suburban, and rural regions in the extreme severity scenario.
\textit{Kaada} is omitted as it does not restore disrupted \acp{RU}.
The \ac{DFR} approach prioritizes urban regions, recovering 62\% of users and 80\% of \acp{RU}, while achieving limited recovery in rural areas (3\% of \acp{RU}).
In contrast, our \ac{SAA}-based solution achieves a more balanced recovery, restoring 62\% of \acp{RU} in urban regions, 14\% in suburban regions, and 24\% in rural regions, with similar behavior when considering user recovery.
While both approaches present higher number of recovered \acp{RU} and users in the urban region, due to its greater density, our proposed \ac{SAA} solution presents a more balanced recovery, with higher number of recovered \acp{RU} in suburban and rural regions.
This difference arises from limited coverage overlap in rural areas, which restricts user re-association after failures.
By accounting for uncertainty in user locations and wireless channel conditions, our solution distributes resources more evenly, whereas deterministic approaches concentrate recovery in high-density regions.
These results highlight the importance of region-aware and multi-metric recovery strategies to improve fairness and overall service restoration.

%% file: Sections/8-conclusion.tex
\section{Conclusion and Future Work}
\label{sec:conclusion}

In this work, we introduced the first adaptive resilience mechanism for disaggregated mobile networks that reinstantiates \ac{RAN} functions by accounting for uncertainty in users' wireless channel conditions and locations during the \textit{in-failure} state.
We evaluate our solution on a real-world mobile network topology, demonstrating its ability to mitigate the impact of cascading failures while achieving higher performance than traditional resilience mechanisms across multiple failure severity levels and traffic demand scenarios.

Our proposed formulation offers many opportunities for future work: \1 investigation of adaptive recovery strategies that account for dynamic traffic variations, enabling resilience mechanisms to better respond to heterogeneous and time-varying network conditions;
\2 exploration of the temporal dimension of the recovery process by designing adaptive failure detection and resilience mechanisms that optimize both recovery performance and the time to recover; and
\3 consideration of multi-dimensional utility metrics beyond throughput, including latency and service-level requirements, to better capture application-aware resilience.

%% file: Sections/acknowledgment.tex
\section{Acknowledgment}
The research leading to this paper received support from the Commonwealth Cyber Initiative, an investment in the advancement of cyber R\&D, innovation, and workforce development. For more information, visit: \url{www.cyberinitiative.org}.
This work also received support from the National Science Foundation US-Ireland R\&D Partnership program under grant No. 2421362 (Resilient Networks project).
This work was supported by CAPES, by MCTIC/CGI.br/FAPESP under grant no. 2020/05127-2 (SAMURAI project), by CNPq under grant no. 306283/2025-5, by RNP/MCTIC under grant no. 01245.020548/2021-07 (Brasil 6G project), and by the OpenRAN Brazil project under grant A01245.014203/2021-14.